\documentclass[fleqn,usenatbib]{mnras}

\usepackage{newtxtext,newtxmath}
\usepackage{multirow}
\usepackage{setspace}
\usepackage[T1]{fontenc}

\DeclareRobustCommand{\VAN}[3]{#2}
\let\VANthebibliography\thebibliography
\def\thebibliography{\DeclareRobustCommand{\VAN}[3]{##3}\VANthebibliography}

\usepackage{graphicx}	% Including figure files
\usepackage{amsmath}	% Advanced maths commands
\usepackage{tikz,hyperref}
\definecolor{lime}{HTML}{A6CE39}
\DeclareRobustCommand{\orcidicon}{
	\begin{tikzpicture}
	\draw[lime, fill=lime] (0,0) 
	circle [radius=0.13] 
	node[white] {{\fontfamily{qag}\selectfont \tiny ID}};
	\draw[white, fill=white] (-0.0625,0.095) 
	circle [radius=0.007];
	\end{tikzpicture}
	\hspace{-2mm}
}

\foreach \x in {A, ..., Z}{\expandafter\xdef\csname orcid\x\endcsname{\noexpand\href{https://orcid.org/\csname orcidauthor\x\endcsname}
			{\noexpand\orcidicon}}
}
\title[6.4 keV line on RS CVn-type stars]{Systematic \textit{NICER} study of the low-ionized Fe K$\alpha$ line on RS Canum Venaticorum type stars}

\author[S. Inoue et al.]{Shun Inoue$^{1}$\orcidA \thanks{E-mail: inoue@cr.scphys.kyoto-u.ac.jp},
Teruaki Enoto$^{1,2}$\orcidB,
Yuta Notsu$^{3,4}$\orcidC,
Hiroyuki Uchida$^{1}$\orcidD,
Wataru Buz Iwakiri$^{5}$\orcidE,
\newauthor  
Kosuke Namekata$^{1,6,7,8}$\orcidF,
and Keith Gendreau$^{9}$\orcidI
\\
$^{1}$Department of Physics, Kyoto University, Kitashirakawa-Oiwake-cho, Sakyo-ku, Kyoto, 606-8502, Japan \\
$^{2}$RIKEN Center for Advanced Photonics (RAP), 2-1 Hirosawa, Wako, Saitama 351-0198, Japan \\
$^{3}$Laboratory for Atmospheric and Space Physics, University of Colorado Boulder, 3665 Discovery Drive, Boulder, CO 80303, USA\\
$^{4}$National Solar Observatory, 3665 Discovery Drive, Boulder, CO 80303, USA \\
$^{5}$ International Center for Hadron Astrophysics, Chiba University, Inage-ku, Chiba, 263-8522, Japan \\
$^{6}$ The Hakubi Center for Advanced Research, Kyoto University, Kyoto 606-8302, Japan \\
$^{7}$ Division of Science, National Astronomical Observatory of Japan, NINS, Osawa, Mitaka, Tokyo, 181-8588, Japan \\
$^{8}$ Heliophysics Science Division, NASA's Goddard Space Flight Center, 8800 Greenbelt Road, Greenbelt, MD 20771, USA \\
$^{9}$ Astrophysics Science Division, NASA’s Goddard Space Flight Center, 8800 Greenbelt Road, Greenbelt, MD 20771, USA
}

\date{Accepted 2025 June 25; Received 2025 June 23; in original form 2024 October 17}

\pubyear{2025}

\begin{document}
\label{firstpage}
\pagerange{\pageref{firstpage}--\pageref{lastpage}}
\maketitle


\begin{abstract}
The Fe K$\alpha$ fluorescence line ($\sim 6.4$ keV) has been observed during solar and stellar flares. 
Two emission mechanisms of the Fe K$\alpha$ line, photoionization and collisional ionization, have been discussed, and the aim of this work is to collect evidences for each mechanism employing a statistical correlation approach between the Fe K$\alpha$ line flux and rough flare properties.
Here, we systematically searched the \textit{NICER} (0.2$-$12 keV) archive data for the Fe K$\alpha$ line of RS Canum Venaticorum type stars. 
Among our analyzed 255 observation IDs with a total exposure of $\sim 700$ ks, we found 25 data sets (total $\sim 40$ ks) exhibiting the Fe K$\alpha$ emission line at 6.37$-$6.54 keV with its equivalent width of 44.3$-$578.4 eV: 18 observations during flares, 6 observations during unconfirmed possible flare candidates and one at a quiescent phase.
These observations indicate a positive correlation between the Fe K$\alpha$ line intensity ($L_{\mathrm{K \alpha}}$) and the 7.11$-$20 keV thermal plasma luminosity ($L_{\mathrm{HXR}}$) with its powerlaw index of $0.86 \pm 0.46$ (i.e., $L_{\mathrm{K \alpha}} \propto L_{\mathrm{HXR}}^{0.86\pm0.46}$). 
This correlation in the range of the thermal plasma luminosity $10^{29-33}$ erg s$^{-1}$ is consistent with the photoionization origin of the line.
On the other hand, the equivalent width of the Fe K$\alpha$ line ($\mathrm{EW}_{\mathrm{K \alpha}}$) has a negative correlation with the 7.11$-$20 keV thermal plasma luminosity with its powerlaw index of $-0.27 \pm 0.10$ (i.e., $\mathrm{EW}_{\mathrm{K \alpha}} \propto L_{\mathrm{HXR}}^{-0.27\pm0.10}$). 
This anti-correlation is consistent with the decline of the fluorescence efficiency with increasing the stellar flare loop height.
Furthermore, we found a signature of an absorption line at $6.38^{+0.03}_{-0.04}$ keV during a superflare of $\sigma$ Gem. The equivalent width of the line was $-34.7^{+2.03}_{-1.58}$ eV.
We discuss the density of the Fe ions from the equivalent width using the curve of growth analysis.
\end{abstract}

\begin{keywords}
X-rays: stars -- stars: late-type -- stars: coronae -- stars: flare 
\end{keywords}

%%%%%%%%%%%%%%%%%%%%%%%%%%%%%%%%%%%%%%%%%%%%%%%%%%
%%%%%%%%%%%%%%%%% BODY OF PAPER %%%%%%%%%%%%%%%%%%
% \input{./table/stars}

% \begin{figure*}
%  \includegraphics[width=10.5cm]{./figure/UX_Ari/lc/lc_U3_paper.pdf}
%  \vspace{0.6cm} \\
%  \includegraphics[width=16.5cm]{./figure/UX_Ari/spec/spec_U3_paper.pdf}
%  \caption{(a) 64 s binned 0.3$-$4 keV count rate (cps; $\mathrm{count \: s^{-1}}$) of UX Ari during Flare U3. The one standard deviation statistical error bars are smaller than the symbol size. The red shaded area indicates the time when the low-ionized Fe emission line is detected. (b)(c) Background-subtracted 5$-$8 keV \textit{NICER} spectra of UX Ari duing GTI 0$-$2 of ObsID 1100380108 fitted with a CIE model (\texttt{apec}; panel b) and  CIE model with the additional Gaussian at $\sim 6.4$ keV (\texttt{apec+gauss}; panel c).   All best-fit parameters of this spectrum are summarized in Table \ref{tab:1100380108_block012}.}
%  \label{fig:UX_Ari_U3}
% \end{figure*}

% \renewcommand{\arraystretch}{1.3}
% \input{./table/UX_Ari/1100380108_block012}
% \renewcommand{\arraystretch}{1.0}

%TC:ignore
 
\section{Introduction}
The low-ionized Fe K$\alpha$ fluorescence line ($\sim 6.4$ keV) has been observed during stellar flares on pre-main sequence \citep{Imanishi_2001, Favata_2005, Tsujimoto_2005, Giardino_2007, Czesla_2010, Stelzer_2011, Pillitteri_2019, Vievering_2019, Castro_2024} and RS CVn and M-type (late type) stars \citep{Osten_2007, Testa_2008, Osten_2010, Huenemoerder_2010, Karmakar_2017}.
The origin of this line has been interpreted as irradiation of the surrounding disk by the stellar X-ray emission in the former case \citep[``disk origin'';][]{Tsujimoto_2005}.
In the latter case, there is a widely accepted view that the low-ionized iron ions in the photosphere are excited by coronal X-rays \citep[``photosphere origin'';][]{Testa_2008, Drake_2008}.
The hydrodynamic modeling of flare loops enabled us to discuss the inclination angle and position of the flare loop by associating the equivalent width of the line to the irradiated area of the photosphere \citep{Testa_2007, Testa_2008, Kowalski_2024}.
Both of these two emission mechanisms of the Fe K$\alpha$ line on pre-main sequence and late type stars are X-ray photoionization.
In addition to this, some studies have considered the contribution of the collisional ionization by nonthermal electrons to the Fe K$\alpha$ line additionally \citep{Osten_2007, Osten_2010}.

Solar flares also show the Fe K$\alpha$ emission line \citep{Neupert_1967, Neupert_1971, Doschek_1971, Feldman_1980, Culhane_1981, Parmar_1984, Tanaka_1984, Tanaka_1985, Emslie_1986, Zarro_1992}.
Similar to stellar flares, photoionization and collisional ionization have been discussed as the emission mechanism of solar flares \citep[e.g.,][]{Bai_1979, Zarro_1992}.
\citet{Bai_1979} indicated the photoionizing flux to generate the Fe K$\alpha$ emission line, $F_{\mathrm{K \alpha}}$, is calculated as 
\begin{equation}
    F_{\mathrm{K \alpha}} = \frac{\Gamma (T, h) f (\theta) }{4 \pi d^{2}} \int_{7.11 \mathrm{keV}}^{\infty} L (\epsilon) d\epsilon,
    \label{eq:photoionization}
\end{equation}
where $\Gamma (T, h)$ is the fluorescence efficiency, depending on the plasma temperature $T$ and the loop height of the flare $h$, $f (\theta)$ the angular dependence of the fluorescence, $d$ the distance to the observer (i.e., between the Earth and Sun), and the integral term the total luminosity above Fe K edge energy ($\sim 7.11$ keV).

If photoionization is the dominant process to produce the stellar Fe K$\alpha$ line, we would expect the Fe K$\alpha$ line luminosity to be proportional to the total luminosity above Fe K edge energy from Equation \ref{eq:photoionization} (see \cite{Pillitteri_2019} for a similar discussion of the Class I star Elias 29).
However, no study has been made confirming the correlation between the Fe K$\alpha$ line luminosity and the total luminosity above Fe K edge energy due to the lack for enough number of observations of the Fe K$\alpha$ line during stellar flares on late type stars.

The good photon statistic is essential to detect the stellar Fe K$\alpha$ line.
RS CVn-type stars are magnetically active and frequently produce large superflares exceeding $10^{35}$ erg \citep[e.g.,][]{Patkos_1981, Rodono_1986, Rodono_1987, Walter_1987, Tsuru_1989, Doyle_1991, Mathioudakis_1992, Kuerster_1996, Endl_1997, Gudel_1999, Osten_1999,Osten_2000, Franciosini_2001, Gudel_2002, Osten_2003, Osten_2004, Brown_2006, Osten_2007, Pandey_2012, Tsuboi_2016, Sasaki_2021, Kawai_2022, Inoue_2023, Karmakar_2023, Kurihara_2024, Inoue_2024b, Gunther_2024, Cao_2024a, Cao_2024b, Didel_2025, Cao_2025}.
\citet{Osten_2007} reported the 6.4 keV emission line during a large superflare, releasing $6 \times 10^{36}$ erg in 0.01$-$200 keV, on the RS CVn-type star II Peg.
Therefore, RS CVn-type stars are the best targets thanks to its high luminosity during a flare.

In this study, we searched \textit{Neutron Star Interior Composition ExploreR} \citep[\textit{NICER};][]{2016SPIE.9905E..1HG} archival data of RS CVn-type stars for the low-ionized Fe K$\alpha$ emission line and investigated their properties.
Though \textit{NICER} is well suited for the detection of the Fe K$\alpha$ line thanks to its large effective area \citep[600 $\mathrm{cm}^{2}$ at 6 keV;][]{Gendreau_2012, Arzoumanian_2014}, there has been no report of the line observed by \textit{NICER} so far.
We describe data reduction and analyses (Section \ref{sec:reduction_analyses}), results (Section \ref{sec:results}), discussion (Section \ref{sec:discussion}), and conclusion (Section \ref{sec:conclusion}). 
In this paper, we use the chi-squared statistics to analyze spetra and 90 \% confidence level as reported error ranges unless otherwise indicated.

\begin{table}
	\centering
	\caption{The distance to RS CVn-type stars in this study. \\ References: (1) \citet{Hummel_2017} (2) \citet{Gaia_2016} (3) \citet{Gaia_2018} (4) \citet{Sasaki_2021} (5) \citet{Roettenbacher_2015} (6) \citet{Pasham_2022} (7) \citet{Bailer-Jones_2021} (8) \citet{Gaia_2021}}
	\label{tab:stars}
\begin{tabular}{crr}
\hline 
Star         &  $d$          & Reference\\
             &  (pc)         &           \\ \hline
UX Ari       &  52 & (1) \\
GT Mus       &  110 &    (2), (3), (4)      \\
$\sigma$ Gem &  39 & (5) \\
HD 251108     &  505  & (6), (7)    \\
HR 1099      &  29 &  (8)  \\
VY Ari       &  42 &  (8) \\       
DS Tuc       &  44 &  (8)  \\   
\hline 
\end{tabular}
\end{table}
\begin{table*}
	\centering
	\caption{List of observations with detections of the low-ionized Fe K$\alpha$ line. $E_{l}$, $F_{\mathrm{K \alpha}}$, $L_{\mathrm{K\alpha}}$, and $\mathrm{EW_{\mathrm{K\alpha}}}$ are the line center energy, photon flux, luminosity, and equivalent width of the Fe K$\alpha$ line, respectively. $L_{\mathrm{HXR}}$ means the thermal luminosity in the 7.11$-$20 keV band. Note that the flare with $\dag$ signs is an unconfirmed possible flare candidate. For the equivalent width with the $*$ signs, the errors are not shown because they could not be calculated.}
	\label{tab:detection_parameters_list}
\begin{tabular}{clccrrrrr}
\hline 
Star                          & Flare     & Obs-ID     & GTI   & $E_{l}$ & $F_{\mathrm{K \alpha}}$ & $L_{\mathrm{K \alpha}}$ & EW$_{\mathrm{K \alpha}}$ & $L_{\mathrm{HXR}}$ \\
                              &           &            &       & (keV)   & ($10^{-4}$ photons cm$^{-2}$ s$^{-1}$)&  ($10^{30}$ erg s$^{-1}$)          & (eV)                     & ($10^{30}$ erg s$^{-1}$)      \\ \hline
\multirow{7}{*}{UX Ari}       & U1$^{\dag}$ & 1100380101 & 0     & $6.48 \pm 0.07$ & $0.98 \pm 0.59$ &$0.31 \pm 0.18$ & $398.9 \pm 307.6$ & $3.25  \pm 2.37$ \\
                              & U2$^{\dag}$ & 1100380106 & all   & $6.53 \pm 0.04$ & $1.38 \pm 0.88$ &$0.43 \pm 0.27$ & $57.2  \pm 36.1$  & $30.91 \pm 4.89$ \\
                              & U3          & 1100380108 & 0+1+2 & $6.43 \pm 0.02$ & $1.77 \pm 0.45$ &$0.55 \pm 0.14$ & $72.7  \pm 19.9$  & $35.49 \pm 1.74$ \\
                              & U4          & 1100380109 & all   & $6.37 \pm 0.32$ & $0.68 \pm 0.59$ &$0.21 \pm 0.18$ & $92.1  \pm 79.8$  & $7.28  \pm 2.59$ \\
                              & U4          & 1100380113 & all   & $6.43 \pm 0.07$ & $0.18 \pm 0.15$ &$0.06 \pm 0.05$ & $187.8^{*}$   & $0.13  \pm 0.10$ \\
                              & U4          & 1100380118 & all   & $6.34 \pm 0.08$ & $0.29 \pm 0.17$ &$0.09 \pm 0.05$ & $578.4 \pm 344.0$ & $0.10  \pm 0.03$ \\
                              & U5$^{\dag}$ & 1100380127 & 3+4   & $6.51 \pm 0.15$ & $0.82 \pm 0.74$ &$0.26 \pm 0.23$ & $259.6^{*}$   & $< 2.46$ \\      \hline
\multirow{2}{*}{GT Mus}       & G1          & 1100140102 & 5     & $6.50 \pm 0.09$ & $3.57 \pm 2.41$ &$5.25 \pm 3.54$ & $377.3 \pm 286.3$ & $60.08 \pm 38.21$ \\    
                              & Quiescent    & 1100140108 & all   & $6.44 \pm 0.06$ & $0.28 \pm 0.19$ &$0.42 \pm 0.28$ & $160.0 \pm 107.6$ & $4.84  \pm 2.37$  \\ \hline
\multirow{3}{*}{$\sigma$ Gem} & S1           & 1200040104 & 0     & $6.38 \pm 0.04$ & $-1.98 \pm -0.87$ &$-0.36 \pm 0.16$ & $-34.4 \pm 1.8$ & $33.17 \pm 3.58$ \\
                              & S1           & 1200040104 & 5     & $6.45 \pm 0.10$ & $1.97 \pm 1.76$ &$0.35 \pm 0.32$ & $107.2 \pm 104.1$ & $6.50 \pm 3.18$ \\
                              & S1           & 1200040106 & all   & $6.49 \pm 0.07$ & $0.36 \pm 0.28$ &$0.07 \pm 0.05$ & $48.1 \pm 34.9$ & $3.17 \pm 0.54$ \\     \hline
HD 251108                     & HD1          & 5203530103 & all   & $6.45 \pm 0.07$ & $0.39 \pm 0.30$ &$12.31 \pm 9.37$ & $87.4 \pm 70.8$ & $431.66 \pm 170.68$        \\ \hline 
\multirow{13}{*}{HR1099}      & HR1$^{\dag}$ & 1114010117 & 5     & $6.47 \pm 0.16$ & $0.83 \pm 0.77$ & $0.09 \pm 0.08$ & $257.3^{*}$ & $< 0.98$ \\
                              & HR2          & 1114010120 & 7     & $6.39 \pm 0.06$ & $1.13 \pm 0.94$ & $0.12 \pm 0.10$ & $48.5 \pm 45.3$ & $4.34 \pm 1.16$ \\
                              & HR2          & 1114010121 & 5     & $6.40 \pm 0.08$ & $0.71 \pm 0.51$ & $0.07 \pm 0.05$ & $81.5 \pm 64.1$ & $1.73 \pm 0.51$ \\
                              & HR2          & 1114010122 & 10+11 & $6.40 \pm 0.05$ & $0.52 \pm 0.34$ & $0.05 \pm 0.04$ & $69.2 \pm 45.5$ & $1.63 \pm 0.37$ \\
                              & HR2          & 1114010123 & 1+2+3 & $6.49 \pm 0.07$ & $0.52 \pm 0.28$ & $0.05 \pm 0.03$ & $111.0 \pm 64.1$ & $0.67 \pm 0.14$ \\
                              & HR3          & 1114010127 & all   & $6.49 \pm 0.10$ & $0.20 \pm 0.14$ & $0.02 \pm 0.01$ & $206.2^{*}$ & $0.07 \pm 0.05$ \\
                              & HR4$^{\dag}$ & 1114010128 & 2     & $6.48 \pm 0.11$ & $0.70 \pm 0.67$ & $0.07 \pm 0.07$ & $95.7 \pm 85.8$ & $1.2 \pm 0.49$ \\
                              & HR4$^{\dag}$ & 1114010128 & 6     & $6.48 \pm 0.08$ & $0.85 \pm 0.64$ & $0.09 \pm 0.07$ & $165.1 \pm 132.6$ & $0.82 \pm 0.5$ \\
                              & HR5       & 1114010133 & 1     & $6.42 \pm 0.06$ & $2.01 \pm 1.27$ & $0.21 \pm 0.13$ & $60.1 \pm 39.9$ & $11.72 \pm 2.75$ \\
                              & HR5       & 1114010133 & 6     & $6.54 \pm 0.04$ & $5.02 \pm 2.41$ & $0.52 \pm 0.25$ & $129.2 \pm 61.8$ & $11.72 \pm 2.85$ \\
                              & HR5       & 1114010133 & 11    & $6.50 \pm 0.06$ & $1.21 \pm 0.96$ & $0.12 \pm 0.10$ & $72.1 \pm 62.9$ & $3.41 \pm 0.87$ \\
                              & HR6       & 1114010136 & 2     & $6.37 \pm 0.04$ & $0.96 \pm 0.55$ & $0.10 \pm 0.06$ & $107.1 \pm 64.3$ & $1.76 \pm 0.54$ \\
                              & HR7       & 1114010153 & 2+3   & $6.38 \pm 0.03$ & $2.26 \pm 1.19$ & $0.23 \pm 0.12$ & $44.3 \pm 23.3$ & $20.83 \pm 1.8$ \\ \hline
\end{tabular}
\end{table*}

\section{Data Reduction \& Analyses}\label{sec:reduction_analyses}

We downloaded \textit{NICER} data of 7 RS CVn-type stars (UX Ari, GT Mus, $\sigma$ Gem, HD 251108, HR 1099, VY Ari, and DS Tuc) from the HEASARC archive.
Table \ref{tab:stars} summarizes the distance to the stars.
% We did not include the data of IM Peg (NICER ObsIDs: 6203900101$-$6203900111), where the Fe XXV He$\alpha$ line was blue-shifted \citep{Inoue_2024b}.
The total number of the data is 255 Obs-IDs with a total exposure of $\sim 700$ ks. 
The data were processed and calibrated in the same manner as that in \citet{Inoue_2024b} with \texttt{nicerl2} in HEASoft ver. 6.32.1 \citep{heasoft_2014} and the calibration database (\texttt{CALDB}) version \texttt{xti20240206}.
We used two filtering criteria for \texttt{nicerl2}: (a) overshoot count rate range of 0$-$5; (b) cut off rigidity greater than 1.5 $\mathrm{GeV \: c^{-1}}.$
After that, we extracted light curves from the filtered event files with \texttt{xselect} and generated ObsID-averaged source and background spectra with \texttt{nicerl3-spect} and the 3C50 model \citep{Remillard_2022}. 
We also extracted time-resolved spectra at each good time interval (GTI) of all Obs-IDs with \texttt{nimaketime}, \texttt{niextract-event} and \texttt{nicerl3-spect}.
We numbered GTIs of each Obs-ID as GTI 0, 1, 2, … by time.
These reduction processes were automated by our publicly available Python script\footnote{\url{https://github.com/seasons0607/NICER_automated_extraction_tool}}.

The quiescent phase of each star is defined as the ObsID that has the lowest count rate among the observations of the target without any apparent variability (Table \ref{tab:quiescent_list} in Appendix \ref{sec:obs_list}).
Then, we calculated the median value ($C_{\mathrm{q}}$) and standard deviation ($\sigma_{\mathrm{q}}$) of the count rates in the 64-sec binned 0.3$-$4 keV light curves of the quiescent phase ObsIDs.
The flare phases are defined as ObsIDs that show average count rates higher than the $1.65 \sigma_{\mathrm{q}}$ of the quiescent rate (i.e., $C_{\mathrm{q}}+1.65\sigma_{\mathrm{q}}$), assuming that the flare exceeds the 90\% confidence interval of the observed temporal fluctuation of the quiescent flux.

To detect the low-ionized Fe K$\alpha$ line, we analyzed both ObsID-averaged and GTI-divided 5$-$8 keV spectra of each ObsID with \texttt{Xspec ver. 12.12.1} \citep{1996ASPC..101...17A} and \texttt{PyXspec ver. 2.1.0} \citep{Gordon_2021}.
While the ObsID-averaged spectra have advantage of the photon statistics, GTI-divided spectra are suited to detect the transiently appearing line emission.
We fitted these 5$-$8 keV spectra with a collisionally-ionized equilibrium (CIE) model (\texttt{apec}) and a CIE with the additional Gaussian component at $\sim 6.4$ keV model (\texttt{apec+gauss}).
We fixed the redshift of \texttt{apec} to 0 because all stars in our sample are galactic sources and the redshift of them can be approximated to 0.

We also fixed the width of the Fe K$\alpha$ gaussian line (\texttt{gauss}) to 0 because the rotational velocity of our analyzed stars is $100$ km s$^{-1}$ $\lesssim$, corresponding to the line width $\sim 5$ eV, which is much smaller than the energy resolution of \textit{NICER} (137 eV at 6 keV\footnote{\url{https://heasarc.gsfc.nasa.gov/docs/nicer/nicer_tech_desc.html}}).
Furthermore, the thermal doppler line broadening for the Fe K$\alpha$ line,
\begin{eqnarray}
    \Delta E_{\mathrm{K\alpha}} = \frac{E_{l}}{c}  \sqrt{\frac{2 k_{B} T_{\mathrm{pho}}}{m_{\mathrm{Fe}}}} \sim 10^{-2} \: \mathrm{eV},
\end{eqnarray}
is also much smaller than the energy resolution of \textit{NICER}, where $E_{l} \sim 6.4$ keV is the line center energy, $c \sim 3 \times 10^{10}$ cm s$^{-1}$ is the light speed, $k_{B} \sim 10^{-16}$ erg K$^{-1}$ is the Boltzmann constant, $T_{\mathrm{pho}} \sim 10^{4}$ K is the temperature of the photosphere, and $m_{\mathrm{Fe}} \sim 10^{-22}$ g is the mass of the iron atom.
In other words, we assumed the \texttt{gauss} was broadened only by the energy resolution of the detector.

We leave the Gaussian line center ($E_{l}$) and normalization ($K^{\mathrm{gauss}}$) free.
When $E_{l}$ comes in the $6.4-6.7$ keV with the 90\% lower limit of $K^{\mathrm{gauss}}$ exceeding zero, the Fe K$\alpha$ emission line is regarded to be detected in the spectrum.
For the special case of IM Peg, where the Fe XXV He$\alpha$ line was blue-shifted \citep{Inoue_2024b}, we need to leave the redshift of \texttt{apec} free to reproduce the blue-shifted Fe XXV He$\alpha$ line. 
In such a case, we have considerable uncertainty regarding the velocity of the low-ionized Fe ions (i.e., whether the Fe K$\alpha$ line is blue-shifted or not); therefore, we did not include the data of IM Peg (\textit{NICER} ObsIDs: 6203900101$-$6203900111) in this study.

When the Fe K$\alpha$ line was detected, we calculated the equivalent width ($\mathrm{EW}_{\mathrm{K \alpha}}$) of it with \texttt{eqwidth} command in \texttt{Xspec} with an option \texttt{"range 0"} to avoid the contamination of the continuum by the strong Fe XXV emission line at $\sim 6.7$ keV \citep{Giardino_2007, Giardino_2009}.
We also calculated the flux of the thermal plasma at 7.11$-$20 keV ($L_{\mathrm{HXR}}$) using the \texttt{flux} command in \texttt{Xspec} since the photons with the higher energy than the Fe K edge at 7.11 keV are captured by the electron in the K shell and induce the low-ionized Fe lines \citep[e.g.,][]{Storm_1970}.

In addition to the narrow-band (5$-$8 keV) fitting with the simple model to focus on the Fe K$\alpha$ line, we also fitted spectra which have the Fe K$\alpha$ line with the two-temperature CIE model with the gauss at $\sim 6.4$ keV (\texttt{vapec+vpec+gauss}) convolved with interstellar absorption (\texttt{tbabs}) in 0.3$-$8 keV to check whether the best-fit parameters of the Fe K$\alpha$ line is significantly affected by the choice of the fitting energy band.
We fixed the redshift and width of the \texttt{gauss} to 0.

Here we evaluate $L_{\mathrm{HXR}}$ as the sum of the emission from the flare and rest of the corona in both narrow- (5$-$8 keV) and wide-band (0.3$-$8 keV) analyses.
However, the emission from the higher temperature component in the wide-band fitting is dominant above 5 keV and corresponding to the one component in the narrow-band fitting.
Then, we consider that the temperature of the higher component in the wide-band fitting and one component in the narrow-band fitting represents the flare plasma.

It should also be pointed out that we set abundances as free parameters in the above narrow- and wide-band fittings for the purpose of accurately estimating the Fe K$\alpha$ line flux.
Accurate modeling of the physical properties of each individual flare is outside the scope of this work, and these limitations are thus acceptable within the defined scope of this work.

\begin{figure*}
 \includegraphics[width=10.5cm]{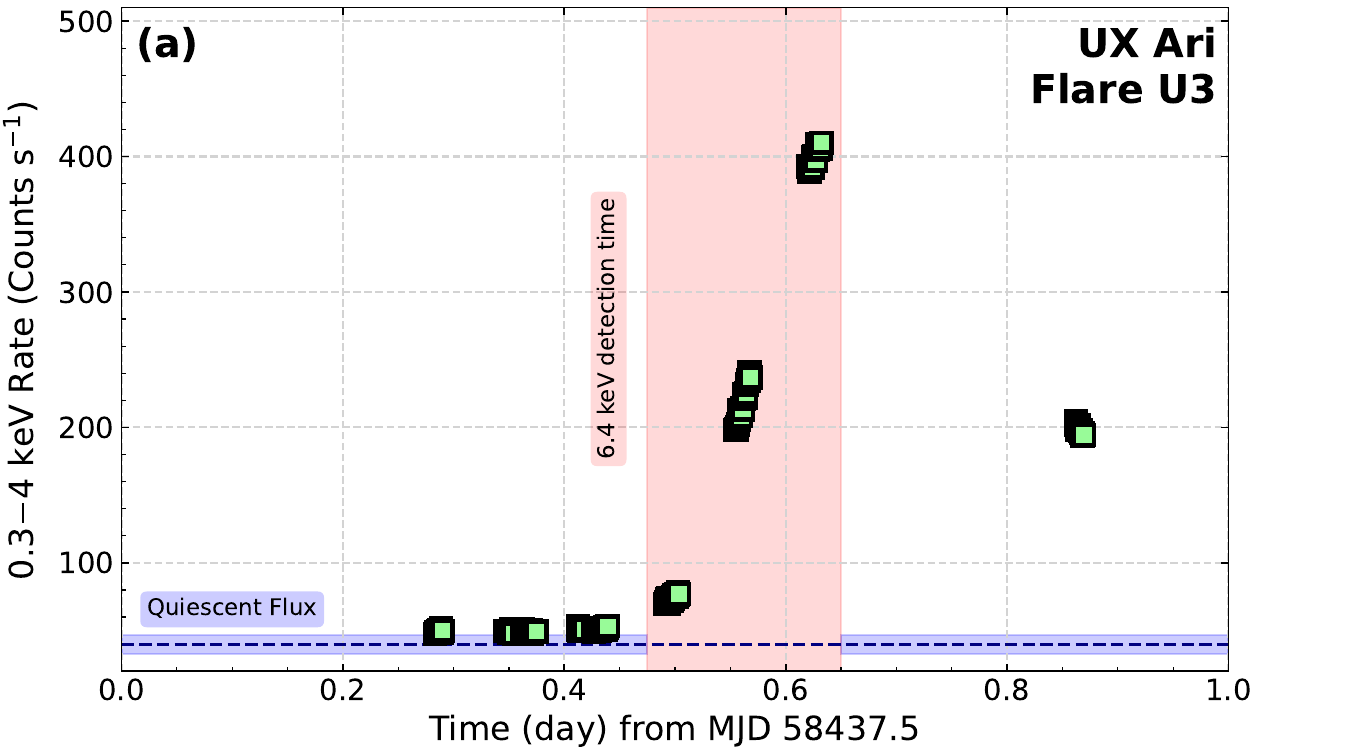}
 \vspace{0.6cm} \\
 \includegraphics[width=16.5cm]{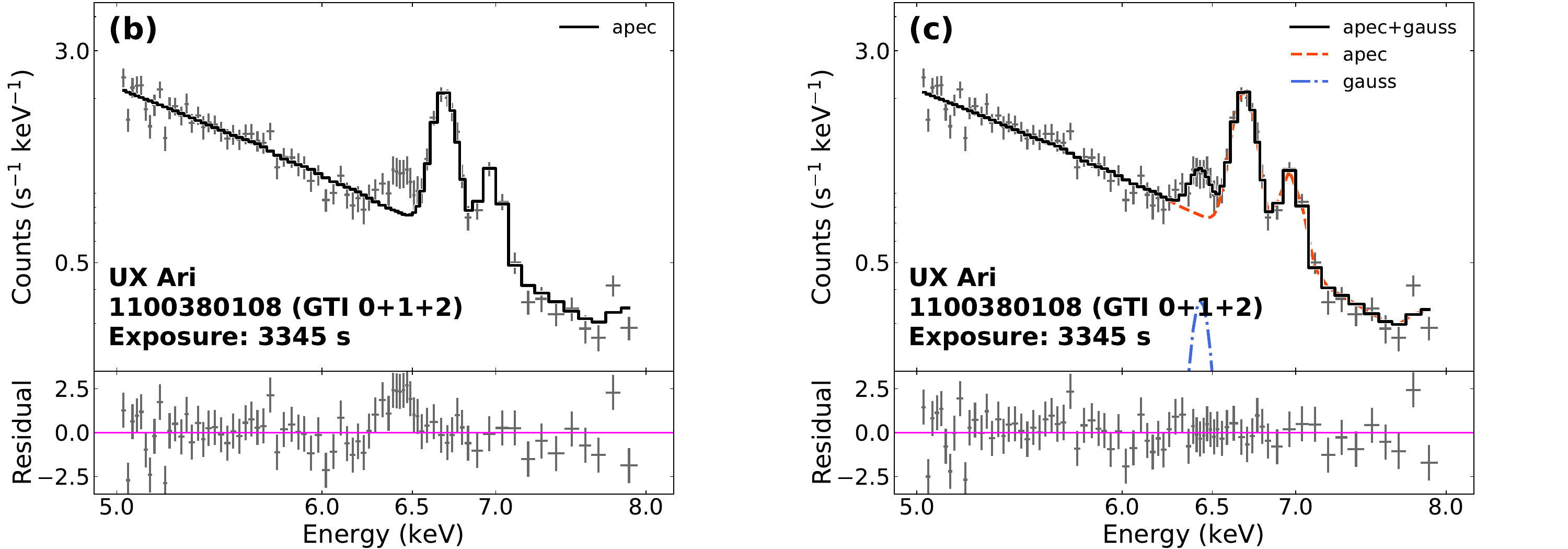}
 \caption{(a) 64 s-binned 0.3$-$4 keV count rate of UX Ari during the flare U3. The one standard deviation statistical error bars are smaller than the symbol size. The red shaded area indicates the time when the low-ionized Fe emission line is detected. (b)(c) Background-subtracted 5$-$8 keV \textit{NICER} spectra of UX Ari duing GTI 0$-$2 of ObsID 1100380108 fitted with a CIE model (\texttt{apec}; panel b) and additionaly with the Gaussian emission line at $\sim 6.4$ keV (\texttt{apec+gauss}; panel c). The spectrum was extracted from the time interval shown in panel a as the red shaded area. All best-fit parameters of this spectrum are summarized in Table \ref{tab:1100380108_block012}.}
 \label{fig:UX_Ari_U3}
\end{figure*}

\renewcommand{\arraystretch}{1.3}
\begin{table*}
\caption{Best-fit spectral parameters of \texttt{apec} and \texttt{apec+gauss} models shown in Figure \ref{fig:UX_Ari_U3}b \& c.}
\begin{center}
\begin{tabular}{cccccc}
\hline 
\multicolumn{6}{c}{1100380108 GTI 0$+$1$+$2 (During Flare U3 on UX Ari)}  \\ \hline 
\multicolumn{3}{c|}{Without the additional Gaussiann}  & \multicolumn{3}{c}{With the additional Gaussiann} \\ \hline
\multirow{3}{*}{\texttt{apec}}                          & $kT$ (keV) / $T$ (MK)        & \multicolumn{1}{c|}{$7.6^{+0.40}_{-0.46}$ / $88.2^{+4.7}_{-5.3}$} & \multirow{3}{*}{\texttt{apec}}    & $kT$ (keV) / $T$ (MK)        & $7.5^{+0.44}_{-0.49}$ /  $87.0^{+5.2}_{-5.7}$   \\
                                                        & $v \: (\mathrm{km \: s^{-1}})$               & \multicolumn{1}{c|}{$0.00$ (fix)} &                                   & $v \: (\mathrm{km \: s^{-1}})$ & $0.00$ (fix)   \\
                                                        & $K^{\mathrm{apec}}$              & \multicolumn{1}{c|}{$0.21^{+0.01}_{-0.01}$} &                                   & $K^{\mathrm{apec}}$ & $0.21^{+0.01}_{-0.01}$    \\ \cline{4-6}
\multirow{3}{*}{---}                          & --- & --- & \multirow{3}{*}{\texttt{gauss}}    & $E_{l}$ (keV) &   $6.43^{+0.02}_{-0.02}$ \\
                                                    & --- & --- &                                  & $\sigma$(keV) &   $0.00$ (fix)  \\
                                                    & --- & ---  &                       & $K^{\mathrm{gauss}}$ ($10^{-4}$) &  $1.77^{+0.45}_{-0.45}$  \\ \hline
\multicolumn{2}{c}{$\chi^{2}$ (d.o.f.)}                                     & \multicolumn{1}{c|}{111 (72)} & \multicolumn{2}{c}{$\chi^{2}$ (d.o.f.)} & 70 (70) \\
\multicolumn{2}{c}{Null hyp. prob.} & \multicolumn{1}{c|}{0.002} & \multicolumn{2}{c}{Null hyp. prob.} &   0.48  \\ \hline 
\end{tabular}
\end{center}
\label{tab:1100380108_block012}
\end{table*}

\renewcommand{\arraystretch}{1.0}

\begin{figure}
\centering
 \includegraphics[width=8.8cm]{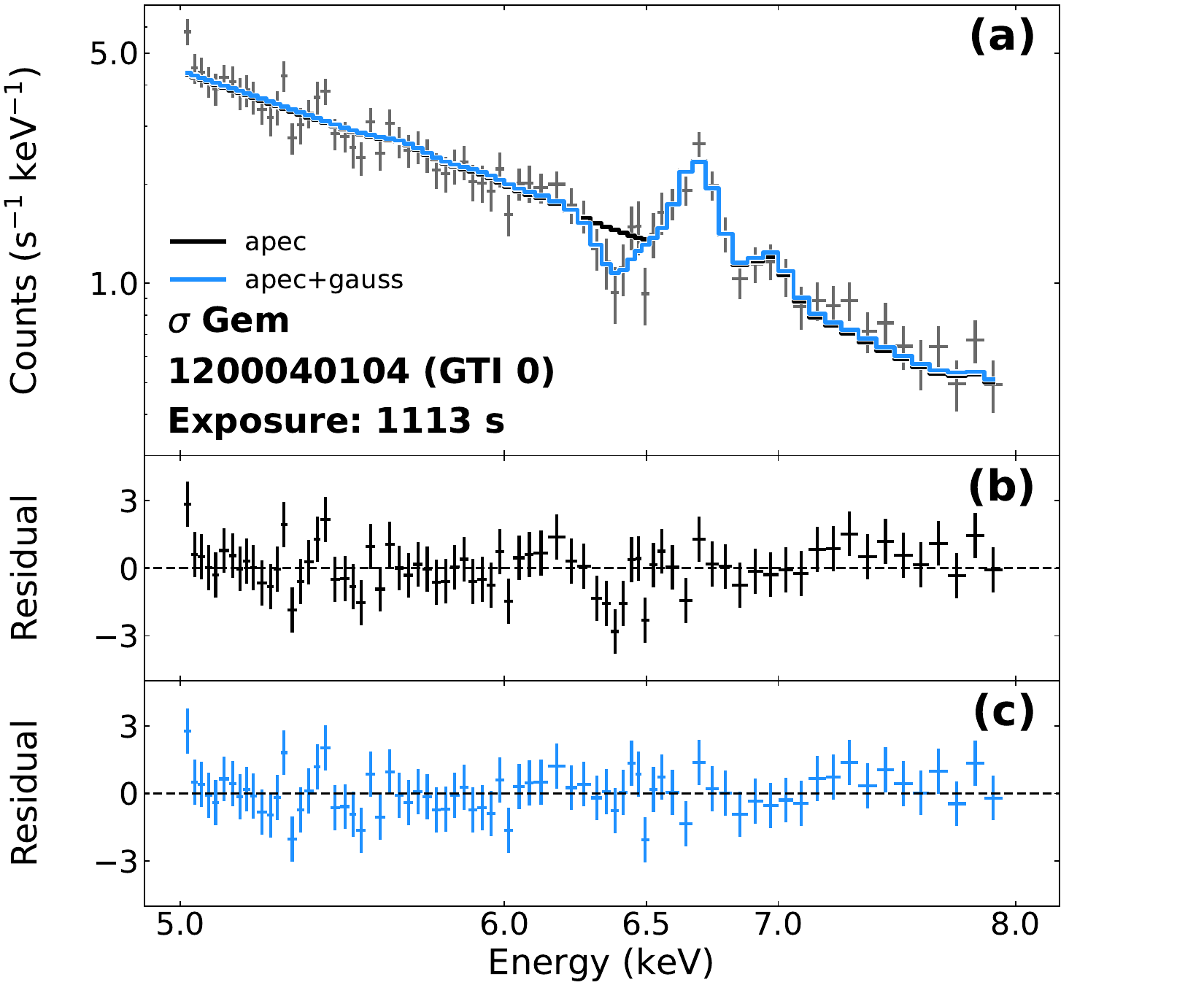}
 \caption{Background-subtracted 5$-$8 keV \textit{NICER} spectrum of $\sigma$ Gem duing the decay phase of Flare S1 (GTI 0 of ObsID 1200040104) fitted with a CIE model (\texttt{apec}; black) and  CIE model with the absorption line at $\sim 6.4$ keV (\texttt{apec+gauss}; blue) with their residuals, shown in panel b and c, respectively. All best-fit parameters of this spectrum are summarized in Table 9 in Online Material.}
 \label{fig:Sigma_Gem_S1_main}
\end{figure}

\begin{figure}
\centering
 \includegraphics[width=8.8cm]{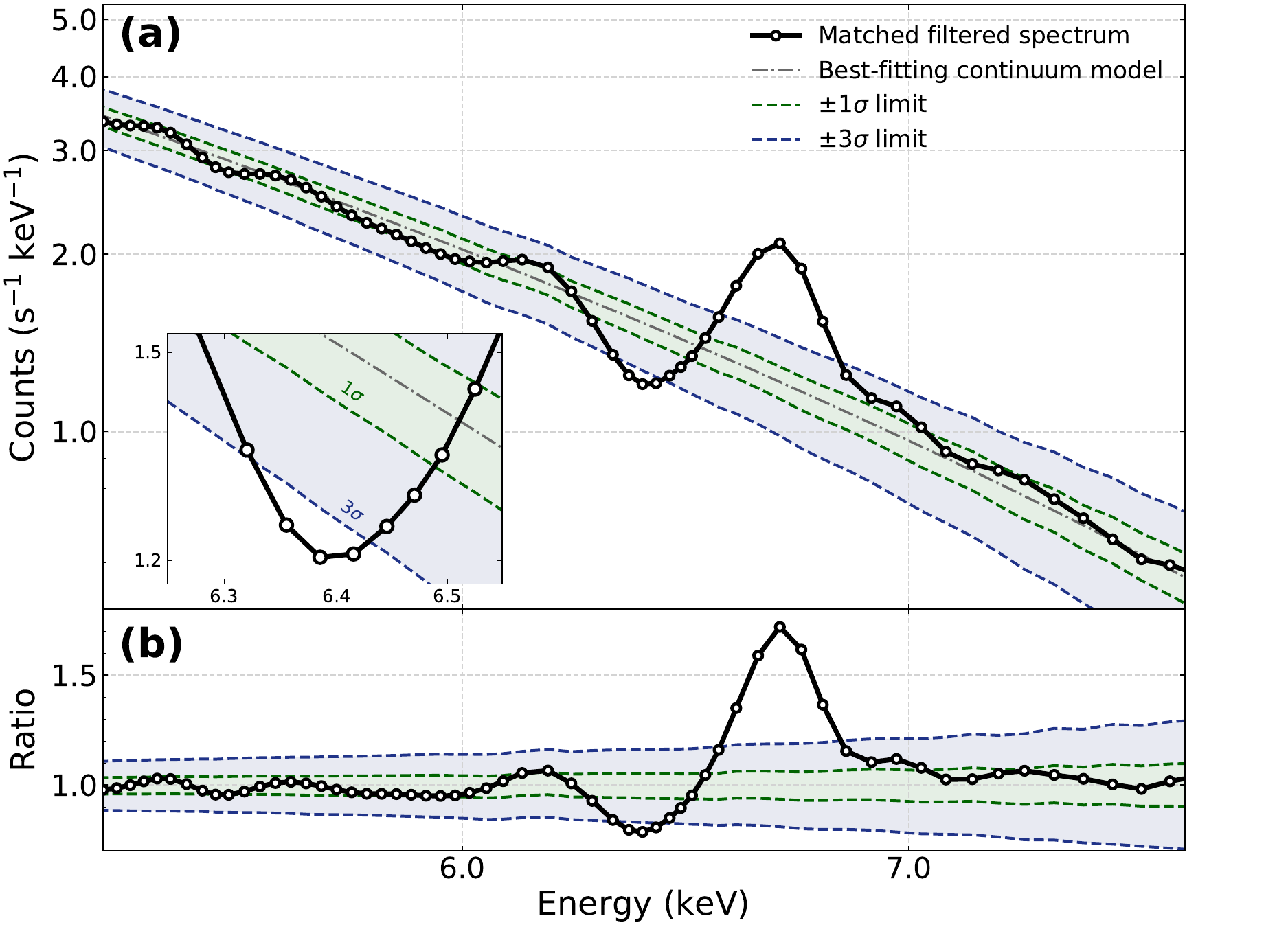}
 \caption{(a) Matched-filtering line search results of the $\sigma$ Gem spectrum (Figure \ref{fig:Sigma_Gem_S1_main}). The solid black line and gray dashdot line indicate the matched-filtered observed spectrum and best-fitting continuum model (\texttt{bremss}), respectively. The green and blue shaded area show the $1 \sigma$ and $3 \sigma$ statistical fluctuations of the continuum model, respectively. 
 The inset panel shows a zoomed view of the region around the absorption feature at $\sim 6.4$ keV. (b) The ratio of the matched filtered spectrum and statistical fluctuations of the continuum model to the best-fit continuum model.}
 \label{fig:Sigma_Gem_S1_MC_simulation}
\end{figure}

\section{Results}\label{sec:results}
\subsection{Detection of Fe K$\alpha$ emission lines}\label{sec:emission}
With the criterion mentioned in Section \ref{sec:reduction_analyses}, we found the 24 samples of the Fe K$\alpha$ emission line during flares and one observation during the quiescent phase  by the narrow-band fitting.
Table \ref{tab:detection_parameters_list} summarizes the list of Fe K$\alpha$ detected observations with the best-fit parameters of the line.
 We also show the wide-band fitting results of the Fe K$\alpha$ line parameters in Appendix \ref{sec:multi}.

We refer to the flares with the Fe K$\alpha$ line of UX Ari as U1$-$U5 in chronological order, that of GT Mus as G1, that of $\sigma$ Gem as S1, that of HD 251108 as HD1, and that of HR 1099 as HR1$-$HR7. 
The flare U3, HR5, and HR6 were observed during its impulsive phase and all other flares were observed during their decay phase.
As an example, the light curve and spectrum of the flare U3 are shown in Figure \ref{fig:UX_Ari_U3} with the fitting results of \texttt{apec} and \texttt{apec+gauss} models (cf. Section \ref{sec:reduction_analyses}).
The detailed best-fit parameters are listed in Table \ref{tab:1100380108_block012}. The light curves, spectra, and best-fit parameters of all the other observations of the Fe K$\alpha$ line (cf. Table \ref{tab:detection_parameters_list}) are in Online Material.
During the decay phase of the flare U4, the Fe K$\alpha$ line was observed three times at different times (Figure 3 in Online Material).
GT Mus showed the Fe K$\alpha$ line during the quiescent phase (Figure 6 in Online Material), which is consistent with the recent result by \citet{Kurihara_2025}.
The flare HR5 of HR 1099 showed the Fe K$\alpha$ line three times at $\sim 0.3$ day intervals (Figure 13 in Online Material).
Some of the flares (Flare U1, U2, U5, HR1, and HR4), marked with $\dagger$ signs in Table \ref{tab:detection_parameters_list}, had count rates higher than during the quiescent phase, but the shape of their light curves made us hesitate to claim that they are flare.
We can not guarantee that these events are simple single flares because they do not show clear time variation during the \textit{NICER} observations.
Their high count rates may reflect the long-term variation of the X-ray luminosity in quiescence.
Whether these events are flares or not, they are included in this study because they show the signatures of the Fe K$\alpha$ lines, but note that they are only unconfirmed possible flare candidates.
The data of VY Ari and DS Tuc are not included in Table \ref{tab:detection_parameters_list}, since they don't have enough photon statistics at $\sim 6.4$ keV to discuss their Fe K$\alpha$ line.

We found the 7 observations with the Fe K$\alpha$ line on UX Ari, 2 observations on GT Mus, 3 observations on $\sigma$ Gem, one observation on HD 251108 and 13 observations on HR 1099 (Table \ref{tab:detection_parameters_list}). 
As a more strict evaluation, we also checked the $3 \sigma$ lower limits of the gaussian normalization of our Fe K$\alpha$ samples. As a result, except for one sample (Flare: HR4, ObsID: 1114010128, Block: 2), the 3$\sigma$ lower limits of the gaussian normalization were above 0 for all other cases.
This means that overall discussions in this paper are not changed even with a more strict detection threshold.
Furthermore, we also fitted the low-count spectra (e.g., Figure 9 in Online Material) with Cash statistics \citep{Cash_1979} and confirmed that there are no significant discrepancy between the results of chi-squared and Cash statistics.

\subsection{Signature of an absorption line during a flare on $\sigma$ Gem}\label{sec:absorption}
Interestingly, we found a signature of an absorption line at $\sim 6.4$ keV during a Flare S1 of $\sigma$ Gem.
Figure \ref{fig:Sigma_Gem_S1_main} shows the spectrum of $\sigma$ Gem on 2019 February 5 02:15:06 $-$ 02:33:39 (GTI 0 of Obs-ID 1200040104) with the exposure time of 1113 sec during the decay phase of the Flare S1 (see also Figure 7 and Table 9 in Online Material).
The best-fit parameters of the line center, absorbed photon flux, and equivalent width of the Gaussian component were $E_{l}=6.38^{+0.03}_{-0.04}$ keV, $K^{\mathrm{gauss}}=-1.99^{+0.88}_{-0.86} \times 10^{-4}$ photons cm$^{-2}$ s$^{-1}$, and $\mathrm{EW}_{\mathrm{K \alpha}}=-34.7^{+2.03}_{-1.58}$ eV, respectively.
The absorption line was observed only during this GTI ($\sim 1000$ sec) and the spectra of the other GTIs and Obs-IDs of $\sigma$ Gem showed no signature of it.
On the other hand, the Fe K$\alpha$ emission line was detected at $\sim 0.5$ and $\sim 2$ days after the observation of the absorption line (Figure 7 in Online Material).

We tested the significance of the absorption line using the matched-filtering line search method \citep{Rutledge_2003, Hurkett_2008, Miyazaki_2016}.
The procedure of our investigation is as follows:
\begin{enumerate}
    \item We fitted the 5$-$8 keV observed spectrum with the bremsstrahlung model (\texttt{bremss}) ignoring the iron line bands (i.e., 5$-$8 keV except for 6.2$-$7.2 keV). Then, we generated $10^{4}$ simulated spectra with the \texttt{fakeit} command in \texttt{Xspec} assuming the parameters of the best-fit bremsstrahlung model.
    \item We processed the $10^{4}$ simulated spectra with the matched filter following the Equation 1 of \citet{Miyazaki_2016}. In this process, we calculated the full width at half-maximum (FWHM) of \textit{NICER} from the Response Matrix File (\texttt{rmf}) file, which was also used to fit the observed spectrum described above.
    \item We also applied the same matched filter to the observed spectrum (Figure \ref{fig:Sigma_Gem_S1_main}) to maximize the signal-to-noise ratio. 
    \item We investigated the statistical distribution of the count rates of the matched filtered fake spectra for each energy bin and calculated the $1 \sigma$ and $3 \sigma$ limits.
\end{enumerate}

Figure \ref{fig:Sigma_Gem_S1_MC_simulation} is the result of the significance study.
The count rates around the line center of the absorption feature come below the $3 \sigma$ significance limit of the continuum.
Thus, from our simulation result, the significance of the absorption line was estimated to be at $\sim 3.8 \sigma$ level.

We also extracted the spectra of the GTI 0 of ObsID 1200040104 for each Measurement Power Unit (MPU) of \textit{NICER} with \texttt{nifpmsel} and \texttt{nicel3-spect}.
We analyzed the MPU-divided spectra to investigate the possibility that a specific MPU caused the absorption line by instrumental reasons. As a result, the 90\% upper limit of the best-fit parameter of the \texttt{gauss} was below 0 for MPU 0, 4, 5, and 6 (Appendix \ref{sec:MPU-divided}). 
The spectra of MPU 1, 2, and 3 also showed possible signatures of the absorption line and the subtle differences among the MPUs should be explained by the statistical fluctuation.
Therefore, we consider that it is highly possible that the astronomical phenomenon made the observed absorption feature.
We discuss the absorption mechanism in Section \ref{sec:absorption_discussion}.

We also detected a similar signature of a P Cygni profile at $\sim 6.4$ keV in the HR 1099 spectrum of the GTI 3 of ObsID 1114010119 (Appendix \ref{sec:PCyg_HR1099}). 
However, we will not discuss this feature in the present paper because the error range of the normalization of \texttt{gauss} was large.

\section{Discussion}\label{sec:discussion}
\subsection{Relationship between the Fe K$\alpha$ line and thermal plasma} \label{sec:correlation}
\begin{figure}
\centering
 \includegraphics[width=8.5cm]{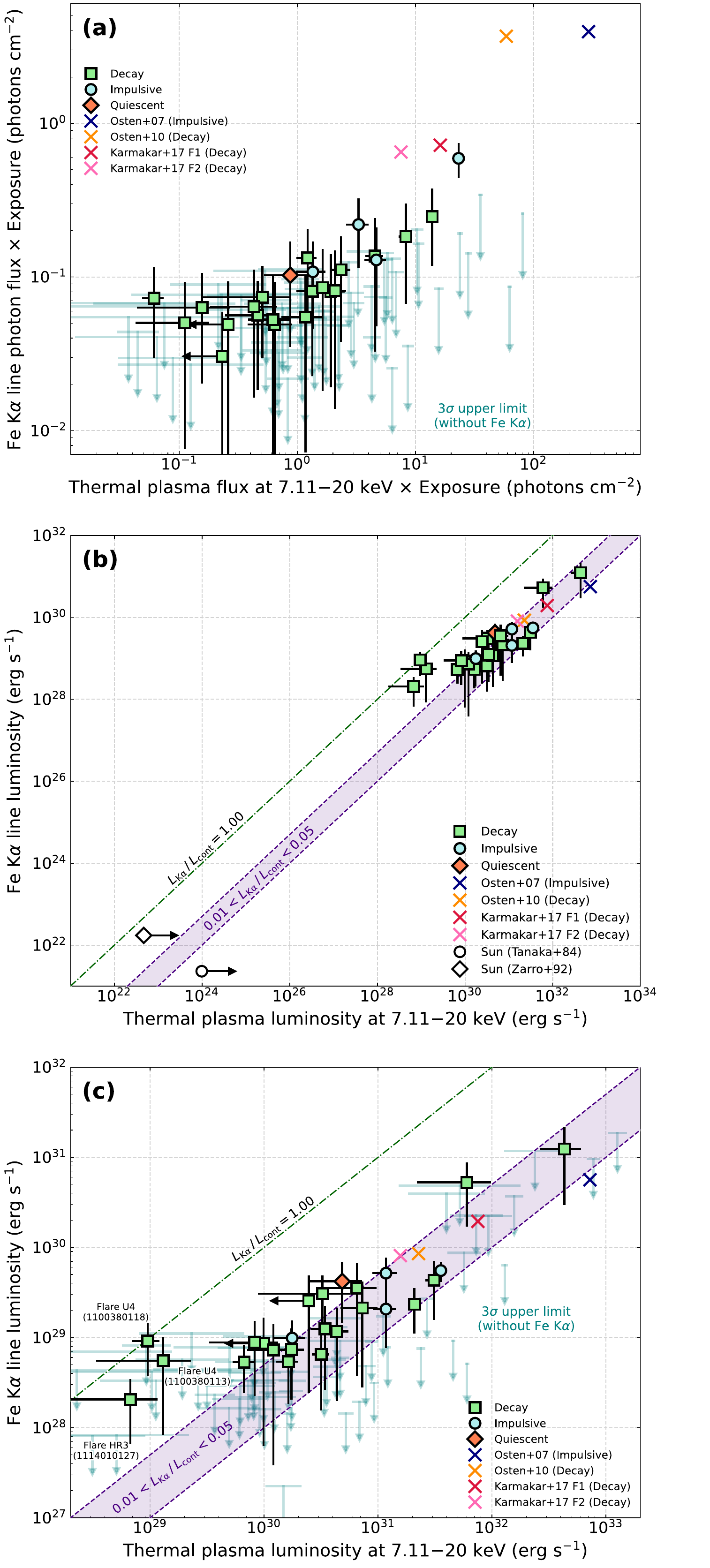}
 \caption{(a) Photon fluence of the Fe K$\alpha$ line and the 7.11$-$20 keV thermal plasma continuum with measure flux multiplied by an exposure. 
 Distance is not corrected.
 Green square, blue circle, and orange diamond symbols are the data obtained by \textit{NICER} during the impulsive phase of a flare, the decay phase of a flare, and the quiescent phase, respectively.
 The crosses represent observations of the stellar flares reported by \citet{Osten_2007, Osten_2010} and \citet{Karmakar_2017}.
The error bars show the 90\% confidence ranges. 
 The green arrows indicate the 3 $\sigma$ upper limits of the Fe K$\alpha$ lines. (b) The Fe K$\alpha$ line luminosity vs. the 7.11$-$20 keV luminosity of the thermal plasma. Distance is corrected (Table \ref{tab:stars}). The black circle and diamond are the solar flares reported by \citet{Tanaka_1984} and \citet{Zarro_1992}, respectively. The purple shaded area shows the range of $L_{\mathrm{K \alpha}} / L_{\mathrm{HXR}}$ from $0.01$ to $0.05$. The green dashdot line shows the slope $L_{\mathrm{K \alpha}} / L_{\mathrm{HXR}} = 1$. (c) Same as panel b but enlarged around the \textit{NICER} detected stellar flares.}
 \label{fig:HXRluminosity_vs_Kaluminosity_with_lit}
\end{figure}

The number of Fe K$\alpha$ detections have demonstrated that \textit{NICER}'s good performance to detect the Fe K$\alpha$ line.
Figure \ref{fig:HXRluminosity_vs_Kaluminosity_with_lit}a shows the relationship between the photon fluence (photons cm$^{-2}$) of the Fe K$\alpha$ line and the thermal plasma. 
In Figure \ref{fig:HXRluminosity_vs_Kaluminosity_with_lit}a, we plotted the 3 $\sigma$ upper limits of the Fe K$\alpha$ line fluence judged by our detection criterion (Section \ref{sec:reduction_analyses}). 
We calculated these upper limits only for the data with enough photon statistic, of which 5--8 keV count rate is higher than the lowest rate among the Fe K$\alpha$ detected sample (e.g., ObsID 1100380118).
Figure \ref{fig:HXRluminosity_vs_Kaluminosity_with_lit}a indicates that this \textit{NICER} study provided the $\sim 10$ times fainter Fe K$\alpha$ line during smaller flares than the previous works \citep{Osten_2007, Osten_2010, Karmakar_2017} conducted by the X-ray telescope \citep[XRT;][]{Burrows_2005} on NASA’s Neil Gehrels Swift Observatory (\textit{Swift}).
This improvement is mainly due to the difference of the effective area: the \textit{NICER} ($600$ cm$^{-2}$) and \textit{Swift} ($100$ cm$^{-2}$) at 6 keV since the detection of the Fe K$\alpha$ line at this flare intensity is mainly dominated by the statistical uncertainties. The photon limit of the Fe K$\alpha$ line is inversely proportional to the effective area. 
The reason for the further improvement of the detection ($\sim 10$ times), compared with the effective area ratio ($\sim 6$ times), can be the sample bias; Previous \textit{Swift} studies \citep{Osten_2007, Osten_2010, Karmakar_2017} only reported the flares at which the Fe K$\alpha$ line was clearly detected.

\begin{figure}
\centering
 \includegraphics[width=8.5cm]{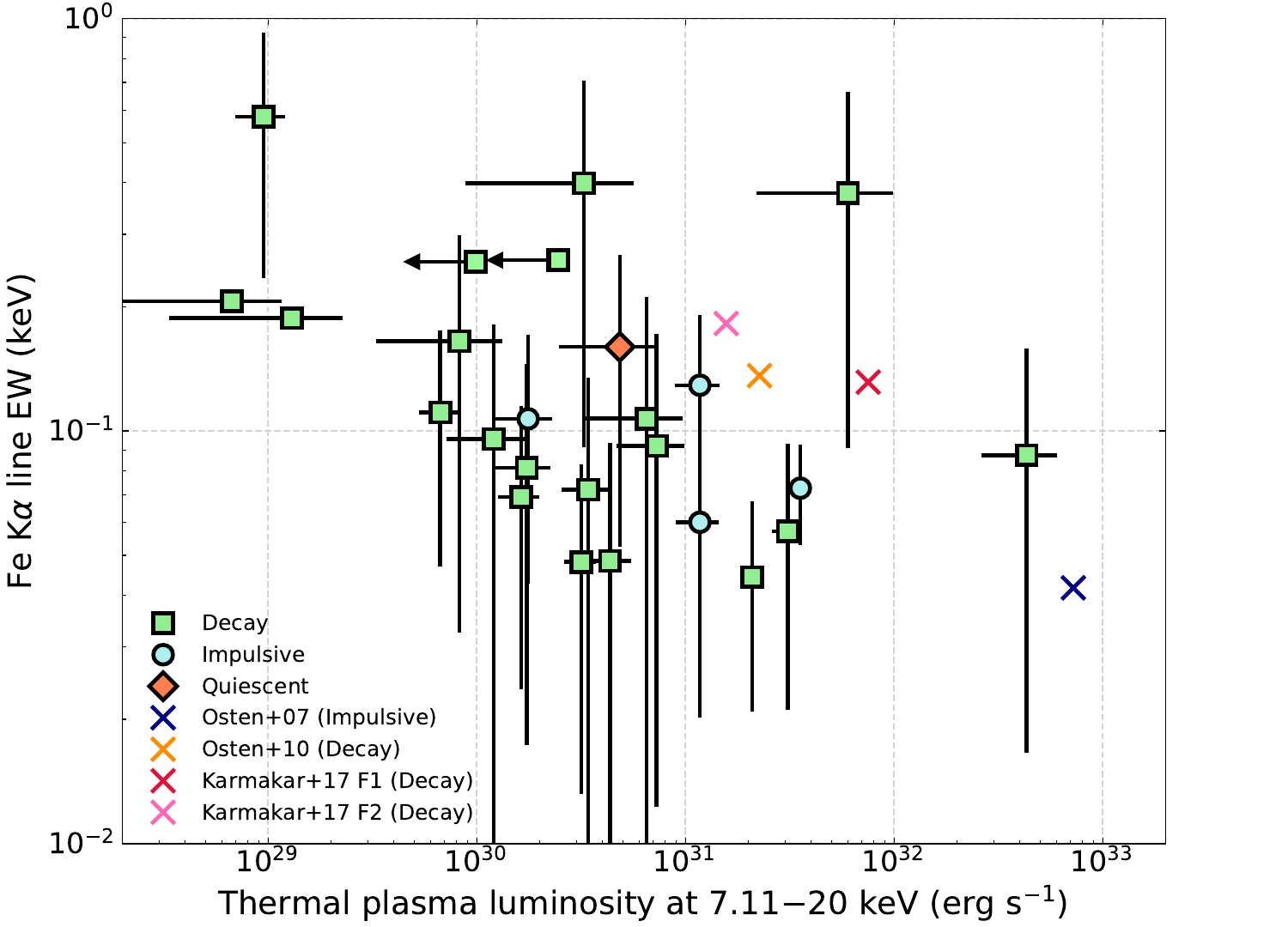}
 \caption{The equivalent width of the Fe K$\alpha$ line vs. the 7.11$-$20 keV luminosity of the thermal plasma. The same symbols as in Figure \ref{fig:HXRluminosity_vs_Kaluminosity_with_lit} are used.}
 \label{fig:HXRluminosity_vs_KaEW_with_lit}
\end{figure}

% The 12 samples of the Fe K$\alpha$ emission line can be used to investigate the emission mechanism of it.
As shown in Figure \ref{fig:HXRluminosity_vs_Kaluminosity_with_lit}b and c, we confirmed the correlation between the $7.11-20$ keV luminosity of thermal plasma ($L_{\mathrm{HXR}}$) and that of the Fe K$\alpha$ emission line ($L_{\mathrm{K \alpha}}$).
We also plotted the data of stellar flares reported by \citet{Osten_2007, Osten_2010} and \citet{Karmakar_2017}, and solar flares reported by \citet{Tanaka_1984} and \citet{Zarro_1992} using their best-fit parameters of the spectra.
We converted the Fe K$\alpha$ line photon flux to its luminosity by multiplying the line energy at 6.4 keV $\sim 1.0 \times 10^{-8}$ erg. 
The calculated $L_{\mathrm{HXR}}$ of \citet{Tanaka_1984} and \citet{Zarro_1992} is regarded to be a lower limit because they only estimated the electron temperature from the line spectra.
Although the 3$\sigma$ upper limits of the Fe K$\alpha$ luminosity for the non-detected sample are not in significant contradiction with the $L_{\mathrm{K \alpha}}-L_{\mathrm{HXR}}$ correlation proposed from the detected sample (Figure \ref{fig:HXRluminosity_vs_Kaluminosity_with_lit}c), future verification by more sensitive observations is needed.
We also investigated the relationship between $\mathrm{EW}_{\mathrm{K \alpha}}$ and $L_{\mathrm{HXR}}$ in Figure \ref{fig:HXRluminosity_vs_KaEW_with_lit}.

Figure \ref{fig:HXRluminosity_vs_Kaluminosity_with_lit}b and c show a clear positive correlation between $L_{\mathrm{K\alpha}}$ and $L_{\mathrm{HXR}}$.
Contrary to the case of the luminosity, the equivalent width of the Fe K$\alpha$ line appears to be negatively correlated with $L_{\mathrm{HXR}}$ (Figure \ref{fig:HXRluminosity_vs_KaEW_with_lit}).
To quantitatively evaluate these correlations between the parameters of thermal plasma and the Fe K$\alpha$ line, we calculated the Spearman rank correlation coefficients ($\mu$).
We also evaluated the standard deviation ($\sigma_{\mu}$) of the correlation coefficients using a simple Montecarlo simulation.
We simulated $10^{5}$ mock samples of the observed data in Figure \ref{fig:HXRluminosity_vs_Kaluminosity_with_lit}c and \ref{fig:HXRluminosity_vs_KaEW_with_lit}, at which each data point randomly generated with following the normal distribution of the observed points in these figures.
Then, we calculated the correlation coefficients for each simulated sample and investigated the distribution of their values. Here, we only include the \textit{NICER} data because it is difficult to estimate the error of previous studies \citep{Osten_2007, Osten_2010, Karmakar_2017}.

\begin{figure}
\centering
 \includegraphics[width=8.5cm]{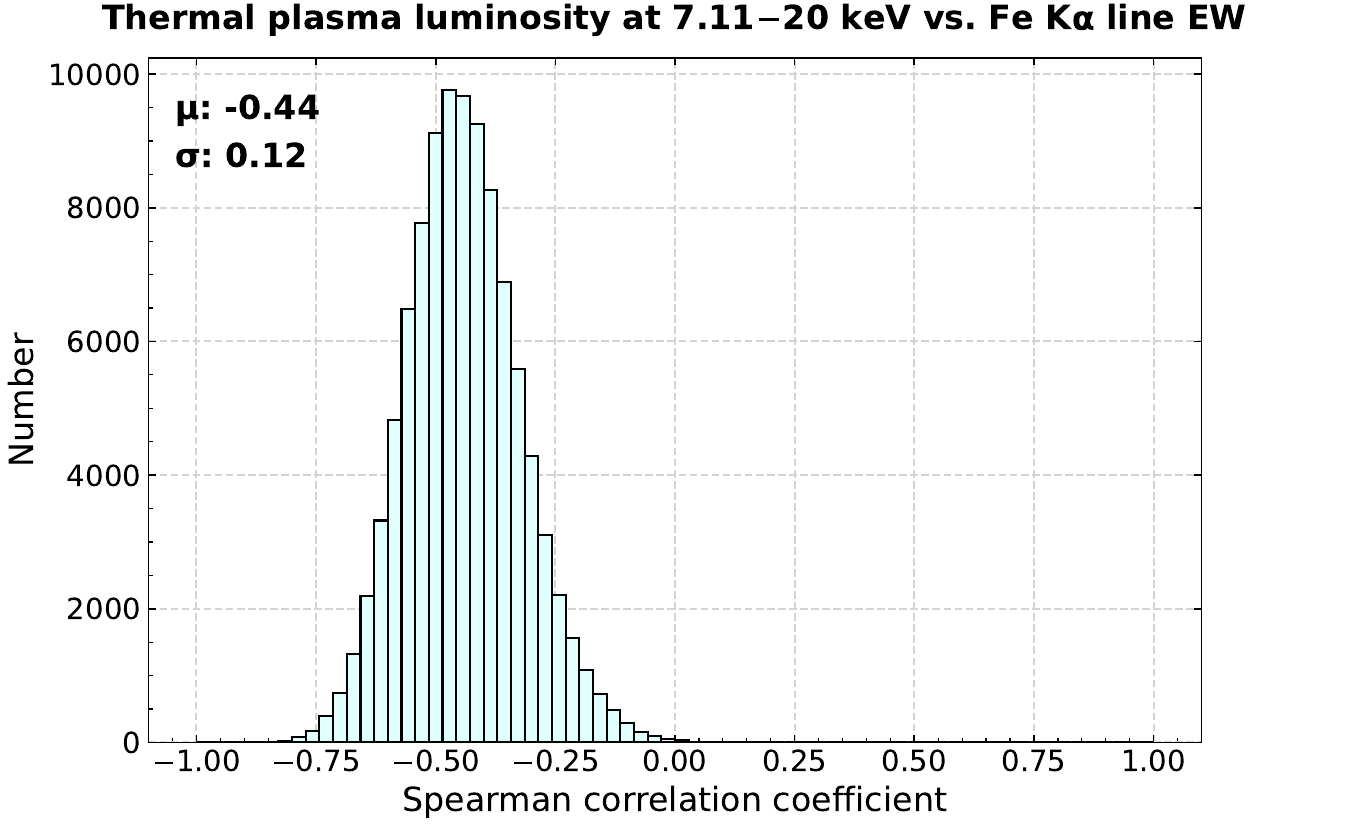}
 \caption{The distribution of the Spearman rank correlation coefficients of the $10^{5}$ mock samples produced from our MonteCarlo simulation between the equivalent width of the Fe K$\alpha$ line and the luminosity of the 7.11$-$20 keV thermal plasma.}
 \label{fig:correlation_HXRluminosity_vs_KaEW_hist}
\end{figure}

As a result, the median value and standard deviation of the correlation coefficient between $L_{\mathrm{K\alpha}}$ and $L_{\mathrm{HXR}}$ are $\mu=0.74$ and $\sigma_{\mu}=0.08$, respectively.
We can confirm the $3\sigma_{\mu}$ lower limit of the correlation coefficient is $0.50 $ $(>0)$. Thus the positive correlation in Figure \ref{fig:HXRluminosity_vs_Kaluminosity_with_lit}b and c above is quantitatively confirmed.
On the other hand, the correlation coefficient between the equivalent width of the Fe K$\alpha$ line and $L_{\mathrm{HXR}}$ is $\mu \pm \sigma_{\mu} =-0.44 \pm 0.12$. 
Figure \ref{fig:correlation_HXRluminosity_vs_KaEW_hist} shows the distribution of the correlation coefficients between $\mathrm{EW_{K\alpha}}$ and $L_{\mathrm{HXR}}$ as an example. 
The $3\sigma_{\mu}$ upper limit of the correlation coefficient between $\mathrm{EW_{K\alpha}}$ and $L_{\mathrm{HXR}}$ is $-0.08$ $(<0)$.
From this, the negative correlation in Figure \ref{fig:HXRluminosity_vs_KaEW_with_lit} is also quantitatively confirmed.
% We also investigated the correlation coefficients between the temperature ($kT$) of the thermal plasma and $L_{\mathrm{HXR}}$ and equivalent width of the Fe K$\alpha$ line.
Table \ref{tab:correlation} summarizes the correlation coefficients and its error among parameters.
% The $3\sigma_{\mu}$ lower limit of the correlation coefficient between $kT$ and $L_{\mathrm{K \alpha}}$ is greater than 0.
% This correlation also supports the photoinonization mechanism in that the spectrum of the hotter plasma is harder.
%On the other hand, we cannot confirm any correlations between $kT$ and $\mathrm{EW}_{\mathrm{K \alpha}}$ quantitatively.

We also calculated the power law index ($\alpha$) and its standard deviation ($\sigma_{\alpha}$) with the same technique (Table \ref{tab:correlation}).
The power-law relations of $L_{\mathrm{K \alpha}}$ and $\mathrm{EW}_{\mathrm{K \alpha}}$ as a function of $L_{\mathrm{HXR}}$ are $L_{\mathrm{K \alpha}} \propto L_{\mathrm{HXR}}^{0.86\pm0.46}$ and $\mathrm{EW}_{\mathrm{K \alpha}} \propto L_{\mathrm{HXR}}^{-0.27\pm0.10}$, respectively.

We also confirmed that these $L_{\mathrm{HXR}}-L_{\mathrm{Ka}}$ correlation and $L_{\mathrm{HXR}}-\mathrm{EW}_{\mathrm{Ka}}$ anti-correlation are obtained by the wide-band and two-temperature spectral analysis.
The correlation coefficients and powerlaw indexes are consistent within the errors between the two fitting method (Appendix \ref{sec:multi}).

\subsection{Emission mechanism of the Fe K$\alpha$ line} \label{sec:emission_mechanism}
There are three possible processes of the Fe K$\alpha$ line:
\begin{enumerate}
    \item Photoionization at the stellar photosphere by hard X-rays above the Fe K edge energy emitted from the thermal plasma in the flare loop \citep[e.g.,][]{Bai_1979}.
    \item Photoionization at the stellar photosphere by hard X-rays emitted when non-thermal electrons accelerated from the reconnection point collide with the footpoint of the flare loop \citep[e.g.,][]{Tanaka_1984}. 
    \item Collisional ionization at the stellar photosphere by the accelerated non-thermal electrons \citep[e.g.,][]{Zarro_1992}.
\end{enumerate}
These processes can occur simultaneously. In this section, we try to interpret the correlation in Section \ref{sec:correlation} by the thermal photoionization process (i), as the dominated process. 

Firstly, as shown in Figure \ref{fig:HXRluminosity_vs_Kaluminosity_with_lit}a, even with enough photon statistic to detect the Fe K$\alpha$ line, $\sim 70$ \%of the data do not show clear Fe K$\alpha$ line emission, only giving upper limits on it. This non-detection can be interpreted by considering the geometry of the flare loop.
The location of the Fe K$\alpha$ line in all processes described above is the photosphere.
Therefore, when a flare occurs on the far side of the star from us, the thermal emission of the flare loop outside the stellar limb can be observed, whereas the Fe K$\alpha$ line from the photosphere would be blocked by the star.

Secondly, the strong correlation between $L_{\mathrm{K\alpha}}$ and $L_{\mathrm{HXR}}$ (Figure \ref{fig:HXRluminosity_vs_Kaluminosity_with_lit}b and c) is consistent with the photoionization by the thermal plasma.
In this process, the Fe K$\alpha$ line luminosity is proportional to the total luminosity above Fe K edge energy (Equation \ref{eq:photoionization}).
On the other hand, this correlation does not necessarily reject other mechanisms because the flares with the large thermal luminosity are thought to have high flux of energetic nonthermal electrons.
Collisional excitation mechanism should also work efficiently during such energetic flares.
In other words, this correlation only indicate the correlation between the flare energy scale and the luminosity of the Fe K$\alpha$ line.

Using Equation \ref{eq:photoionization}, the ratio of the Fe K$\alpha$ line luminosity ($L_{\mathrm{K\alpha}}$) to that of the $7.11-20$ keV thermal plasma ($L_{\mathrm{HXR}}$) is calculated as 
 \begin{equation}
    \frac{L_{\mathrm{K\alpha}}}{L_{\mathrm{HXR}}} = \frac{4 \pi d^{2} F_{\mathrm{K\alpha}}}{L_{\mathrm{HXR}}} \sim \Gamma (T, h) f (\theta).
\end{equation}
If the emission mechanism of the Fe K$\alpha$ line is photoionization only, this ratio should be restricted to
\begin{equation}
    0.01 < \Gamma (T, h) f (\theta) < 0.05,
\end{equation}
assuming the ranges of $\Gamma (T, h)$ and $f (\theta)$ shown in Table 1 and Figure 3 of \citet[][]{Bai_1979}, respectively.
Most data points are in this range as shown in the purple shaded area in Figure \ref{fig:HXRluminosity_vs_Kaluminosity_with_lit}b and c.
Only the data of Flare U4 and HR3 exceed this range outside the error.
This may suggest that not only thermal photoionization but also collisional ionization or photoionization by nonthermal electrons contribute to the radiation of the Fe K$\alpha$ line duting Flare U4 and HR3. 
However, in general, these non-thermal processes occur during the impulsive phase of a flare \citep[e.g.,][]{Shibata_2011}. It is curious that such bright Fe K$\alpha$ line was observed a few days after the flare peak during these flares (Figure 3 and 11 in Online Material).

On the other hand, the Flare U4’s and HR3's deviation are not clearly seen in the $L_{\mathrm{HXR}}-L_{\mathrm{K\alpha}}$ correlation obtained by the wide-band fitting (Appendix \ref{sec:multi}).

Finally, the negative correlation between $\mathrm{EW}_{\mathrm{K\alpha}}$ and $L_{\mathrm{HXR}}$ (Figure \ref{fig:HXRluminosity_vs_KaEW_with_lit}) is also consistent with the photoionization by the thermal plasma in the flare loop. \citet{Testa_2008} and \citet{Ercolano_2008} indicated that the fluorescence efficiency $\varepsilon = 2 L_{\mathrm{K\alpha}} / L_{\mathrm{HXR}}$ and $\mathrm{EW}_{\mathrm{K\alpha}}$ decline with increasing the loop height $h$ due to the $1/h^{2}$ dilution of the flux at stellar surface by using the 3D radiative transfer code MOCASSIN \citep{Ercolano_2003a, Ercolano_2003b,  Ercolano_2005, Ercolano_2008ApJS}.
The negative correlation shown in Figure \ref{fig:HXRluminosity_vs_KaEW_with_lit} does not contradict the fact that more energetic flares with larger $L_{\mathrm{HXR}}$ have larger flare loop sizes and heights.
Since the equivalent width of the Fe K$\alpha$ line also depends on the angle of flare loop inclination with respect to the line of sight \citep[See Figure 2 and 3 of][]{Testa_2008}, we cannot conclude that the result of Figure \ref{fig:HXRluminosity_vs_KaEW_with_lit} must reflect the fluorescence efficiency only.

\begin{table}
\centering
	\caption{The Spearman rank correlation coefficients between the parameters of thermal plasma and the Fe K$\alpha$ line. The $\mu$ and $\sigma$ parameters mean the median value and standard deviation, respectively.}
	\label{tab:correlation}
\begin{tabular}{crr}
\hline
Parameter combination                          & Correlation coefficient & Power law index\\
                                                          & ($\mu \pm \sigma_{\mu}$)   &   ($\alpha \pm \sigma_{\alpha}$) \\ \hline
$L_{\mathrm{HXR}}$ vs. $L_{\mathrm{K \alpha}}$           & $0.74 \pm 0.08$       & $0.86 \pm 0.46$  \\
$L_{\mathrm{HXR}}$ vs. $\mathrm{EW}_{\mathrm{K \alpha}}$ & $-0.44\pm 0.12$       & $-0.27 \pm 0.10$ \\ \hline
\end{tabular}
\end{table}

\begin{figure}
\centering
 \includegraphics[width=8.5cm]{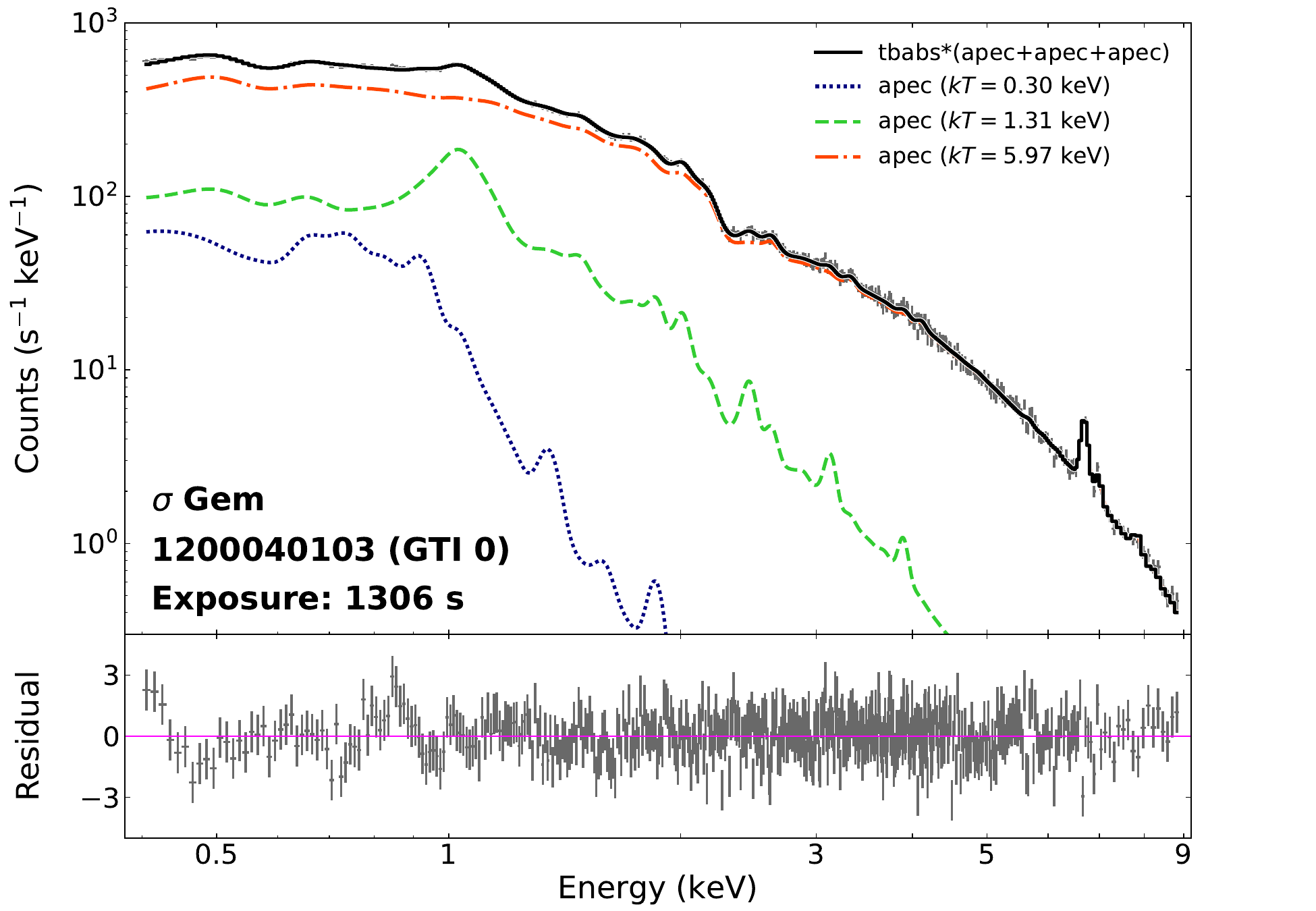}
 \caption{Background-subtracted 0.4$-$9 keV \textit{NICER} spectrum of $\sigma$ Gem duing the decay phase of Flare S1 (GTI 0 of ObsID 1200040103) fitted with the three-temperature CIE model with interstellar absorption (\texttt{tbabs*(vapec+vapec+vapec)}; black). The temperature of the three CIE components are 5.97 keV (orange dashed–dotted), 1.31 keV (green dashed), and 0.30 keV (navy dotted), respectively. All best-fit parameters of this spectrum are summarized in Table \ref{tab:1200040103_block0_wide}.}
 \label{fig:flare_S1_wide}
\end{figure}

\begin{figure*}
\centering
 \includegraphics[width=15.5cm]{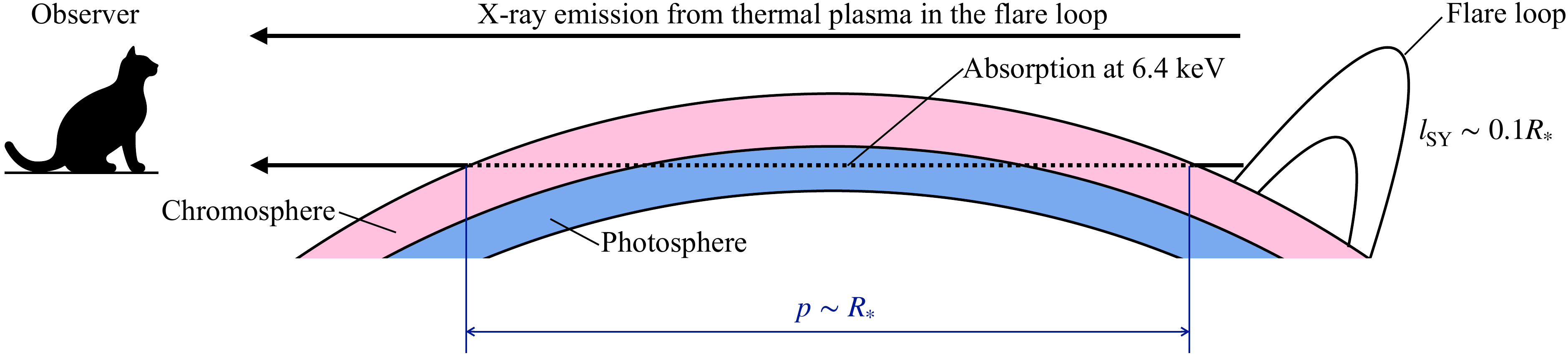}
 \caption{Schematic picture of the loop geometry which makes the Fe K$\alpha$ absorption feature.}
 \label{fig:geometry}
\end{figure*}

\begin{table}
\centering
\caption{Best-fitting parameters of the spectrum of GTI 0 of Obs-ID 1200040103 (Figure \ref{fig:flare_S1_wide}) with three-temperature collisionally ionized models. Norm means $10^{-14} (4 \pi)^{-1} (D_{\mathrm{A}})^{-2} \int n_{\mathrm{e}} n_{\mathrm{H}} dV$, where $D_{\mathrm{A}}$ is the angular diameter distance to the source, $n_{\mathrm{e}}$ and $n_{\mathrm{H}}$ are the electron and hydrogen densities, and $dV$ is the volume element.}
\renewcommand{\arraystretch}{1.1}
\begin{tabular}{lcc}
\hline
\texttt{tbabs}                             & $N_{\mathrm{H}}$ ($10^{20}$ cm$^{-2}$) & 3.31 (fixed) \\ \hline
\multirow{2}{*}{\texttt{vapec} (high temp.)}            & $kT$ (keV)            & $5.97^{+0.19}_{-0.18}$ \\
                                  & norm             & $0.97^{+0.02}_{-0.02}$ \\ \hline
\multirow{2}{*}{\texttt{vapec} (medium temp.)}            & $kT$  (keV)           &  $1.31^{+0.07}_{-0.07}$ \\
                                  & norm             & $0.18^{+0.02}_{-0.02}$  \\ \hline
\multirow{2}{*}{\texttt{vapec} (low temp.)}            & $kT$  (keV)           & $0.30^{+0.02}_{-0.02}$  \\
                                  & norm             &  $0.05^{+0.001}_{-0.001}$\\ \hline
\multicolumn{2}{l}{$\chi^{2}$ (d.o.f)}               &  490 (492) \\ \hline
\end{tabular}
\label{tab:1200040103_block0_wide}
\end{table}
\renewcommand{\arraystretch}{1.0}

\subsection{Absorption of the Fe K$\alpha$ line detected from $\sigma$ Gem}\label{sec:absorption_discussion}
The Fe K$\alpha$ absorption feature was observed at $6.38^{+0.03}_{-0.04}$ keV during the flare S1 of $\sigma$ Gem (Section \ref{sec:absorption}).
Here we discuss the geometry of the flare loop and density of the cool plasma to make the absorption.

We calculated the loop size of the flare S1 using the magnetic reconnection model equation shown in \citet{Shibata_2002},
\begin{eqnarray} 
l_{\mathrm{SY}} = &10^{9}& \left( \frac{EM_{\mathrm{peak}}}{10^{48} \: \mathrm{cm}^{-3}} \right)^{3/5} \nonumber \\ &\times& \left( \frac{\mathnormal{n_{0}}}{10^{9} \: \mathrm{cm}^{-3}} \right)^{-2/5} \left( \frac{\mathnormal{T_{\mathrm{peak}}}}{10^{7} \: \mathrm{K}} \right)^{-8/5}
\: \mathrm{cm}, \label{eq:Shibata_Yokoyama_L}
\end{eqnarray}
where $l_{\mathrm{SY}}$ is the length of the flare loop, $EM_{\mathrm{peak}}$ is the volume emission measure at the flare peak, $T_{\mathrm{peak}}$ is the peak electron temperature, and $n_{0}$ is the preflare coronal density.
In Figure \ref{fig:flare_S1_wide}, we analyzed the spectrum of GTI 0 of Obs-ID 1200040103, which is closest to the peak of the Flare S1, in 0.4$-$9 keV and substituted the best-fit parameters (Table \ref{tab:1200040103_block0_wide}) into Equation \ref{eq:Shibata_Yokoyama_L}. 
We used the three-temperature CIE model with interstellar absorption and linked the abundance among the all components.
The preflare coronal density is assumed to be $n_{0} = 10^{10-13} \: \mathrm{cm^{-3}}$ \citep{Aschwanden_1997, Gudel_2004, Reale_2007, Sasaki_2021} in this calculation. As a result, $l_{\mathrm{SY}}$ is $(0.05-0.8) R_{*}$, where $R_{*} = 10.1 R_{\odot}$ is the radius of the K type star of $\sigma$ Gem \citep{Roettenbacher_2015} and $R_{\odot}$ is the solar radius.

Considering this flare loop size, we speculate the geometry as shown in Figure \ref{fig:geometry}. The flare occurred at the limb of the star.
The absorbers between the flare loop and the observer are the low-ionized Fe ions in the photosphere, chromosphere, and transition region.
The distance ($p$) of the absorption of the thermal X-ray emission is estimated to be the same order of the K-type stellar radius of $\sigma$ Gem ($p \sim R_{*} \sim 10 R_{\odot} \sim 7 \times 10^{11}$ cm).

Curve of growth analysis shows the relationship between the equivalent width of an absorption line and the column density of scattering ions \citep[e.g.,][]{Tombesi_2011}. 
\citet{Kotani_2000} and \citet{Young_2004} studied the curve of growth of the iron ions.
Using the curve of growth of the low-ionized Fe XVIII ions based on its oscillator strength of the K$\alpha$ transition of 0.109 \citep[][]{Behar_2002} shown in Figure 6 of \citet{Young_2004}, we estimate the column density of the low-ionized Fe ion to be $N_{\mathrm{Fe}} \sim 10^{20}$ cm$^{-2}$ taking into account the equivalent width of the Fe K$\alpha$ absorption line ($\mathrm{EW}_{\mathrm{K \alpha}} \sim -35$ eV).
As mentioned in \citet{Young_2004}, Fe ions with lower ionization state than the Fe XVIII ion can not make an absorption line because they do not have an L-shell vacancy. On the other hand, the line center energy of the Fe ions with higher ionization state than the Fe XX ion is clearly higher than the observed value of $6.38^{+0.03}_{-0.04}$ keV \citep[e.g.,][]{Palmeri_2003, Mendoza_2004, Yamaguchi_2014}. Therefore, we used the curve of growth of the Fe XVIII ion \citep{Young_2004}, whose line center energy is the lowest among the Fe ions that can make an absorption line.
The line center energy of the Fe XVIII line is $\sim 6.43$ keV and roughly consistent with the observed value of $6.38^{+0.03}_{-0.04}$ keV.
We also confirmed that the spectral analysis with a fixed line center of \texttt{gauss} to 6.43 keV does not change the reduced $\chi^{2}$ value significantly.

Then, the number density of the low-ionized Fe ion is estimated to be 
\begin{equation}
\label{eq:nfe_SigmaGem}
n_{\mathrm{Fe}} = N_{\mathrm{Fe}} / p \sim 10^{8} \: \mathrm{cm}^{-3}.
\end{equation}
As a comparison, the number density of the iron in the solar photosphere was estimated to be 
\begin{equation}
\label{eq:nfe_Sun}
n_{\mathrm{Fe}} = 10^{\log n_{\mathrm{H}} + A_{\mathrm{Fe}} -12} \: \mathrm{cm^{-3}} \sim 10^{11.5} \: \mathrm{cm^{-3}},
\end{equation}
where $n_{\mathrm{H}} \sim 10^{16}$ cm$^{-3}$ \citep[e.g.,][]{Bommier_2020} is the hydrogen density and $A_{\mathrm{Fe}} \sim 7.5$ \citep[e.g.,][]{Bellot_Rubio_2002} the iron abundance in the solar photosphere.
As shown in Equations \ref{eq:nfe_SigmaGem} and \ref{eq:nfe_Sun}, there is the 3.5th-order difference of the iron density between $\sigma$ Gem and the Sun. 

The discrepancy between $\sigma$ Gem (Equation \ref{eq:nfe_SigmaGem}) and the solar photosphere (Equation \ref{eq:nfe_Sun}) suggests the possibility that the thermal X-ray emission from the flare loop traveled through the less dense plasma, such as the chromosphere or transition layer, rather than the photosphere.
From the chromosphere to the transition layer, the plasma density drops sharply by more than three orders \citep[e.g.,][]{Aschwanden_2004}.
Therefore, there should be a strong density corresponding on the value of Equation \ref{eq:nfe_SigmaGem} in these regions.
In addition, we used the solar photosphere abundance, and accurate Fe abudance of $\sigma$ Gem is needed to be constrained for further discussions. 
For more discussion of the absorption feature at $\sim 6.4$ keV, we need to know the ionization state of the Fe ion with the high-resolution spectroscopy of X-Ray Imaging and Spectroscopy Mission \citep[XRISM;][]{Tashiro_2025} with its microcalorimeter Resolve \citep[][]{Ishisaki_2018} in the future.

\section{Summary and conclusion}\label{sec:conclusion}
We systematically searched the \textit{NICER} archive data of RS CVn-type stars (UX Ari, GT Mus, $\sigma$ Gem, HD 251108, HR 1099, VY Ari, and DS Tuc) for the low-ionized Fe K$\alpha$ line.
Since the number of observations of the line is still very limited on late-type stars it is imperative to increase the sample size of the Fe K$\alpha$ line by taking advantage of the large effective area of \textit{NICER}.
Our main results are as follows:
\begin{enumerate}
    \item We found 25 observations of the Fe K$\alpha$ emission line. The 18 observations were conducted during flares of UX Ari, GT Mus, $\sigma$ Gem, HD 251108, and HR 1099, 6 observations were during unconfirmed possible flare candidates of UX Ari and HR 1099 and another one was during the quiescent phase of GT Mus.
    \item Our 25 spectra indicate a positive correlation between the Fe K$\alpha$ line photon intensity and the 7.11$-$20 keV thermal plasma luminosity with its Spearman rank correlation coefficient of $0.74\pm0.08$ and powerlaw index of $L_{\mathrm{K \alpha}} \propto L_{\mathrm{HXR}}^{0.86\pm0.46}$ ($1 \sigma$ error). They also show the negative correlation between the equivalent width of the Fe K$\alpha$ line and the luminosity of the thermal plasma at 7.11$-$20 keV with its Spearman rank correlation coefficient of $-0.44\pm0.12$ and powerlaw index of $\mathrm{EW}_{\mathrm{K \alpha}} \propto L_{\mathrm{HXR}}^{-0.27\pm0.10}$ ($1 \sigma$ error). These results support the photoionization mechanism of the line.
    \item The Fe K$\alpha$ absorption feature was detected during the decay phase of the $\sigma$ Gem flare on 2019 February. The line center energy, photon intensity, and equivalent width were $6.38^{+0.03}_{-0.04}$ keV, $-1.99^{+0.88}_{-0.86} \times 10^{-4}$ photons cm$^{-2}$ s$^{-1}$, and $-34.7^{+2.03}_{-1.58}$ eV, respectively.
    Using the curve of growth of the Fe XVIII ion \citep{Young_2004}, the observed equivalent width of the absorption line shows the number density of the low-ionized Fe ion estimated to be  $\sim 10^{8} \: \mathrm{cm}^{-3}$.  
\end{enumerate}

We demonstrated that there are many stellar flares showing the Fe K$\alpha$ line and that the photoionization mechanism by the thermal plasma is consistent with the detected data. 
Along with the Fe XXV He$\alpha$ and Fe XXVI Ly$\alpha$ line, the low-ionized Fe K$\alpha$ line becomes a powerful tool not only for the geometry of the flare loop, but also to diagnose the properties of the flare plasma in the XRISM era \citep{Tashiro_2025}.
This \textit{NICER} work is the forerunner of the upcoming XRISM study.

%TC:ignore
\section*{Acknowledgements}
The \textit{NICER} analysis software and data calibration were provided by the NASA \textit{NICER} mission and the Astrophysics Explorers Program.
We sincerely thank the anonymous referee for the helpful comments and suggestions that clearly define the aims/scope of this work.
We thank K. Shibata (Doshisha University) and T. Tsuru (Kyoto University) for their useful comments and discussions.
We thank Z. Arzoumanian and K. Hamaguchi (NASA/GSFC) for their \textit{NICER} operations.
This research is supported by the JSPS KAKENHI grant No. 24KJ1483 (S.I.), 24K00680, and 24H00248 (K.N.).
T.E. was supported by the RIKEN Hakubi project.
Y.N. acknowledge support from NASA ADAP award program No. 80NSSC21K0632.
K.N. was supported by the Hakubi project at Kyoto University

%%%%%%%%%%%%%%%%%%%%%%%%%%%%%%%%%%%%%%%%%%%%%%%%%%
\section*{Data Availability}
All \textit{NICER} data analyzed in this article were retrieved from the publicly available HEASARC archive. 
The processed data are available from the corresponding author S.I. on request.
The csv data file of Figure \ref{fig:HXRluminosity_vs_Kaluminosity_with_lit}c and \ref{fig:HXRluminosity_vs_KaEW_with_lit} can be found in Zenodo (\doi{10.5281/zenodo.15743344}).



\bibliographystyle{mnras}
\bibliography{example} % if your bibtex file is called example.bib

%\begin{thebibliography}{99}
%\end{thebibliography}

%%%%%%%%%%%%%%%%%%%%%%%%%%%%%%%%%%%%%%%%%%%%%%%%%%

%%%%%%%%%%%%%%%%% APPENDICES %%%%%%%%%%%%%%%%%%%%%
\appendix
\section{\textit{NICER} Observation lists}\label{sec:obs_list}
We summarized in Table \ref{tab:quiescent_list} the list of observations which we referred to as the quiescent phase of each star.
We also provide in Table \ref{tab:detection_list} the list of all continuous observations during which the Fe K$\alpha$ line was detected.

\renewcommand{\arraystretch}{1.0}
\begin{table*}
\begin{center}
	\caption{Quiescent phases of our targets. The start and end time are shown in UT. The rates are calculated as the number of events in all energy bands of \textit{NICER} divided by the exposure time.}
	\label{tab:quiescent_list}
\begin{tabular}{crrrrr}
\hline
Stellar Name              & Obs-ID     & Start               &  End                & Exposure    & Rate   \\
                          &            & (Date)              &  (Date)             & (sec)       & (counts s$^{-1}$) \\ \hline
\multirow{4}{*}{UX Ari}   & 1100380102 & 2017-12-06 12:14:52 & 2017-12-06 12:18:20 & 208  & 43.4 \\
                          & 1100380103 & 2017-12-09 05:01:42 & 2017-12-09 23:39:59 & 1146 & 40.7 \\
                          & 1100380104 & 2017-12-10 15:01:02 & 2017-12-10 19:49:20 & 1194 & 41.8 \\
                          & 1100380105 & 2017-12-11 17:44:46 & 2017-12-11 17:46:59 & 133  & 58.0 \\ \hline
\multirow{3}{*}{GT Mus}   & 1100140106 & 2017-11-18 00:27:17 & 2017-11-18 00:32:32 & 315  & 47.9 \\
                          & 1100140107 & 2017-11-19 05:27:46 & 2017-11-19 08:34:24 & 907  & 50.4 \\
                          & 1100140108 & 2017-11-20 00:02:37 & 2017-11-20 18:56:37 & 3602 & 47.2 \\ \hline
$\sigma$ Gem              & 1200040102 & 2018-02-15 16:31:22 & 2018-02-15 18:39:03 & 4066 & 84.7 \\ \hline
\multirow{8}{*}{HD251108} & 5203530161 & 2023-01-21 12:18:41 & 2023-01-21 15:33:40 & 2021 & 4.5 \\ 
                          & 5203530162 & 2023-01-23 02:49:20 & 2023-01-23 02:54:29 & 309  & 5.2 \\
                          & 5203530163 & 2023-01-24 17:38:25 & 2023-01-24 17:53:40 & 915  & 4.0 \\
                          & 5203530164 & 2023-01-25 08:59:16 & 2023-01-25 09:23:00 & 1424 & 4.1 \\
                          & 5203530165 & 2023-01-26 14:24:15 & 2023-01-26 14:41:13 & 1018 & 5.0 \\
                          & 5203530166 & 2023-01-27 05:58:07 & 2023-01-27 10:47:46 & 1329 & 4.6 \\
                          & 5203530167 & 2023-01-29 12:04:13 & 2023-01-29 16:49:40 & 1593 & 3.9 \\
                          & 5203530168 & 2023-02-02 18:41:56 & 2023-02-02 23:33:20 & 1107 & 3.7 \\ \hline
HR1099                    & 1114010113 & 2017-12-16 05:27:06 & 2017-12-16 19:26:03 & 3688 & 60.1\\ \hline
\end{tabular}
\end{center}
\end{table*}
\renewcommand{\arraystretch}{1.0}
\renewcommand{\arraystretch}{1.0}
\begin{table*}
	\centering
	\caption{The observation list of flares during which the Fe K$\alpha$ line is detected. The $*$ and $\dagger$ signs added in the Obs-ID column indicate the data in which the low-ionized Fe K$\alpha$ emission and abosrption line are detected, respectively.}
	\label{tab:detection_list}
\begin{tabular}{ccrrrrr}
\hline 
Stellar Name & \multicolumn{1}{c}{Flare Number} & Obs-ID     & Start               & End                & Exposure    & Rate \\ 
             &                                  &            & (Date)              & (Date)             & (sec)       & (counts s$^{-1}$) \\ \hline
\multirow{16}{*}{UX Ari}     & \multicolumn{1}{c}{U1}    & $^{*}$1100380101 & 2017-11-09 20:46:31 & 2017-11-09 21:18:29 & 1584 & 143.3 \\ \cline{2-7} 
     & \multicolumn{1}{c}{U2}    & $^{*}$1100380106 & 2018-02-22 16:46:51 & 2018-02-22 20:00:20 & 1327 & 443.8 \\ \cline{2-7} 
     & \multirow{2}{*}{U3}       & 1100380107       & 2018-11-15 18:46:43 & 2018-11-15 22:32:20 & 4029 & 53.1  \\
     &                           & $^{*}$1100380108 & 2018-11-15 23:45:55 & 2018-11-16 08:51:20 & 3967 & 235.3 \\ \cline{2-7} 
     & \multirow{11}{*}{U4}      & $^{*}$1100380109 & 2018-11-26 18:57:04 & 2018-11-26 19:10:29 & 805  & 223.8 \\
     &                           & 1100380110       & 2018-11-27 04:10:25 & 2018-11-27 22:58:00 & 3217 & 155.4 \\
     &                           & 1100380111       & 2018-11-28 04:53:04 & 2018-11-28 23:40:20 & 2987 & 109.1 \\
     &                           & 1100380112       & 2018-11-29 05:35:49 & 2018-11-29 21:17:23 & 2726 & 83.9  \\
     &                           & $^{*}$1100380113 & 2018-11-30 03:13:09 & 2018-11-30 22:00:00 & 2841 & 70.1  \\
     &                           & 1100380114       & 2018-12-01 03:55:26 & 2018-12-01 14:59:18 & 2466 & 71.5  \\
     &                           & 1100380115       & 2018-12-02 23:09:45 & 2018-12-02 23:25:20 & 410  & 62.2  \\
     &                           & 1100380116       & 2018-12-03 05:20:26 & 2018-12-03 05:36:00 & 932  & 62.5  \\
     &                           & 1100380117       & 2018-12-04 06:27:05 & 2018-12-04 22:10:20 & 1798 & 58.1  \\
     &                           & $^{*}$1100380118 & 2018-12-05 14:28:27 & 2018-12-05 15:10:20 & 2494 & 56.3  \\
     &                           & 1100380119       & 2018-12-06 12:05:26 & 2018-12-06 12:31:23 & 1557 & 64.2  \\ \cline{2-7} 
     & \multicolumn{1}{c}{U5}    & $^{*}$1100380127 & 2019-01-17 23:55:24 & 2019-01-18 23:15:58 & 2551 & 95.7  \\ \cline{1-7} 
\multirow{5}{*}{GT Mus}     & \multirow{4}{*}{G1}       & 1100140101 & 2017-07-18 17:00:45 & 2017-07-18 23:14:00 & 525 & 314.9 \\
     &                           & $^{*}$1100140102 & 2017-07-19 00:45:15 & 2017-07-19 20:54:20 & 1429 & 258.7 \\
     &                           & 1100140103       & 2017-07-20 03:02:57 & 2017-07-20 20:05:23 & 555  & 193.0 \\
     &                           & 1100140104       & 2017-07-21 06:47:58 & 2017-07-21 08:25:20 & 108  & 150.4 \\ \cline{2-7} 
     & \multicolumn{1}{c}{Quiescent}    & $^{*}$1100140108 & 2017-11-20 00:02:37 & 2017-11-20 18:56:37 & 3602 & 47.2 \\ \cline{1-7} 
\multirow{7}{*}{$\sigma$ Gem}     & \multirow{7}{*}{S1}       & 1200040103 & 2019-02-04 16:58:47 & 2019-02-04 23:42:50 & 5050 & 737.4 \\
     &                           & $^{\dag}$1200040104 & 2019-02-05 02:15:06 & 2019-02-05 22:29:12 & 9677 & 412.7 \\
     &                           & 1200040105          & 2019-02-06 00:01:04 & 2019-02-06 20:15:40 & 4809 & 304.5 \\
     &                           & $^{*}$1200040106    & 2019-02-07 02:08:25 & 2019-02-07 16:24:20 & 3796 & 233.2 \\
     &                           & 1200040107          & 2019-02-08 04:24:26 & 2019-02-08 23:18:59 & 4151 & 200.5 \\
     &                           & 1200040108          & 2019-02-11 03:28:23 & 2019-02-11 03:50:21 & 1311 & 133.4 \\
     &                           & 1200040109          & 2019-02-13 00:17:00 & 2019-02-13 00:33:31 & 985  & 114.7 \\ \cline{1-7} 
\multirow{28}{*}{HD 251108}     & \multirow{28}{*}{HD1}       & 5203530101 & 2022-11-09 18:38:43 & 2022-11-09 23:34:00 & 2185 & 79.6 \\
     &                           & 5203530102          & 2022-11-10 00:51:24 & 2022-11-10 22:49:20 & 6746 & 60.8 \\
     &                           & $^{*}$5203530103    & 2022-11-11 00:05:42 & 2022-11-11 11:13:40 & 2035 & 48.7 \\
     &                           & 5203530104          & 2022-11-12 14:48:42 & 2022-11-12 19:48:00 & 1774 & 33.8 \\
     &                           & 5203530105          & 2022-11-13 04:51:25 & 2022-11-13 05:04:54 & 809  & 30.5 \\
     &                           & 5203530106          & 2022-11-14 01:05:13 & 2022-11-14 14:55:20 & 1221 & 24.9 \\
     &                           & 5203530107          & 2022-11-15 02:01:50 & 2022-11-15 23:55:47 & 5715 & 20.6 \\
     &                           & 5203530108          & 2022-11-16 00:47:26 & 2022-11-16 22:55:17 & 8984 & 19.2 \\
     &                           & 5203530109          & 2022-11-17 00:13:26 & 2022-11-17 20:24:15 & 9024 & 17.3 \\
     &                           & 5203530110          & 2022-11-18 16:25:42 & 2022-11-18 22:43:22 & 2017 & 16.4 \\
     &                           & 5203530111          & 2022-11-19 00:03:53 & 2022-11-19 23:24:22 & 3364 & 16.4 \\
     &                           & 5203530112          & 2022-11-20 00:50:23 & 2022-11-20 22:39:53 & 5629 & 14.6 \\
     &                           & 5203530113          & 2022-11-21 00:07:05 & 2022-11-21 23:52:23 & 4981 & 13.3 \\
     &                           & 5203530114          & 2022-11-22 02:55:40 & 2022-11-22 21:32:33 & 3033 & 12.1 \\
     &                           & 5203530115          & 2022-11-23 02:07:19 & 2022-11-23 15:44:20 & 1394 & 11.7 \\
     &                           & 5203530116          & 2022-11-24 07:02:18 & 2022-11-24 22:34:42 & 2105 & 11.3 \\
     &                           & 5203530117          & 2022-11-25 21:43:00 & 2022-11-25 23:20:41 & 563  & 10.9 \\
     &                           & 5203530118          & 2022-11-26 00:49:02 & 2022-11-26 18:02:13 & 3686 & 11.8 \\
     &                           & 5203530119          & 2022-11-27 23:10:51 & 2022-11-27 23:22:02 & 671  & 9.6 \\
     &                           & 5203530120          & 2022-11-28 00:45:28 & 2022-11-28 16:23:23 & 4693 & 11.3 \\
     &                           & 5203530121          & 2022-11-30 02:25:29 & 2022-11-30 04:07:52 & 1074 & 10.0 \\
     &                           & 5203530122          & 2022-12-01 08:11:39 & 2022-12-01 19:03:29 & 3443 & 9.6 \\
     &                           & 5203530123          & 2022-12-02 00:48:29 & 2022-12-02 13:22:00 & 1390 & 8.6 \\
     &                           & 5203530124          & 2022-12-04 02:50:28 & 2022-12-04 07:32:26 & 557  & 7.8 \\
     &                           & 5203530125          & 2022-12-05 22:04:37 & 2022-12-05 23:45:27 & 661  & 8.5 \\
     &                           & 5203530126          & 2022-12-06 01:15:02 & 2022-12-06 02:53:03 & 487  & 7.6 \\
     &                           & 5203530127          & 2022-12-08 04:05:49 & 2022-12-08 18:08:35 & 1035 & 7.2 \\
     &                           & 5203530128          & 2022-12-09 12:46:17 & 2022-12-09 22:07:20 & 244  & 7.1 \\ \cline{1-7} 
\end{tabular}
    % \footnotemark[$*$]These values are at r1 peak. \\ 
    % \footnotemark[$\dag$]This value is at r2 peak.

\end{table*}
\renewcommand{\arraystretch}{1.0}
\setcounter{table}{1}
\renewcommand{\arraystretch}{1.0}
\begin{table*}
	\centering
	\caption{(Continued.)}
\begin{tabular}{ccrrrrr}
\hline 
Stellar Name & \multicolumn{1}{c}{Flare Number} & Obs-ID     & Start               & End                & Exposure    & Rate \\ 
             &                                  &            & (Date)              & (Date)             & (sec)       & (counts s$^{-1}$) \\ \hline
\multirow{22}{*}{HR 1099}     & \multicolumn{1}{c}{HR1}   & $^{*}$1114010117          & 2017-12-20 08:40:53 & 2017-12-20 23:45:18 & 2321 & 106.8 \\ \cline{2-7} 
     & \multirow{5}{*}{HR2}      & 1114010119          & 2018-02-09 15:33:19 & 2018-02-09 23:31:00 & 4550  & 830.5 \\
     &                           & $^{*}$1114010120    & 2018-02-10 00:50:23 & 2018-02-10 22:45:00 & 12557 & 555.4 \\
     &                           & $^{*}$1114010121    & 2018-02-10 23:58:25 & 2018-02-11 23:22:00 & 15746 & 326.7 \\
     &                           & $^{*}$1114010122    & 2018-02-12 00:38:52 & 2018-02-12 22:32:40 & 16484 & 311.3 \\
     &                           & $^{*}$1114010123    & 2018-02-12 23:46:39 & 2018-02-13 18:38:10 & 7691  & 254.1 \\ \cline{2-7} 
     & \multirow{2}{*}{HR3}      & 1114010126          & 2018-03-01 20:25:00 & 2018-03-01 23:36:53 & 845   & 100.5 \\  
     &                           & $^{*}$1114010127    & 2018-03-02 01:03:02 & 2018-03-02 22:46:13 & 3881  & 90.9  \\ \cline{2-7}
     & \multirow{2}{*}{HR4}      & $^{*}$1114010128    & 2018-07-13 13:59:13 & 2018-07-13 23:27:40 & 5231  & 289.3 \\
     &                           & 1114010129          & 2018-07-14 00:51:37 & 2018-07-14 04:05:40 & 1273  & 274.0 \\ \cline{2-7}
     & \multirow{4}{*}{HR5}      & 1114010132          & 2018-08-16 12:30:16 & 2018-08-16 23:25:20 & 1516  & 188.3 \\
     &                           & $^{*}$1114010133    & 2018-08-17 00:47:15 & 2018-08-17 22:44:59 & 8365  & 570.4 \\
     &                           & 1114010134          & 2018-08-18 00:01:57 & 2018-08-18 23:27:20 & 5257  & 331.4 \\
     &                           & 1114010135          & 2018-08-19 00:46:26 & 2018-08-19 01:00:01 & 813   & 210.7 \\ \cline{2-7}
     & \multirow{7}{*}{HR6}      & $^{*}$1114010136    & 2019-02-09 02:06:49 & 2019-02-09 17:51:14 & 3530  & 177.0 \\
     &                           & 1114010137          & 2019-02-10 01:15:17 & 2019-02-10 18:32:40 & 4842  & 205.8 \\
     &                           & 1114010138          & 2019-02-11 05:02:38 & 2019-02-11 20:50:40 & 3821  & 111.9 \\
     &                           & 1114010139          & 2019-02-12 04:28:50 & 2019-02-12 09:28:00 & 2934  & 87.1  \\
     &                           & 1114010140          & 2019-02-13 05:22:45 & 2019-02-13 20:48:11 & 2219  & 89.0 \\
     &                           & 1114010141          & 2019-02-14 05:57:51 & 2019-02-14 07:34:16 & 459   & 90.4 \\
     &                           & 1114010142          & 2019-02-15 12:28:08 & 2019-02-15 12:44:02 & 954   & 112.1 \\ \cline{2-7}
     & \multicolumn{1}{c}{HR7}   & $^{*}$1114010153    & 2019-03-01 00:58:46 & 2019-03-01 06:15:40 & 2462  & 338.0 \\ \cline{1-7}
\end{tabular}
    % \footnotemark[$*$]These values are at r1 peak. \\ 
    % \footnotemark[$\dag$]This value is at r2 peak.

\end{table*}
\renewcommand{\arraystretch}{1.0}

\section{Wide-band and multi-temperature fit}\label{sec:multi}

\begin{figure}
\centering
 \includegraphics[width=8.5cm]{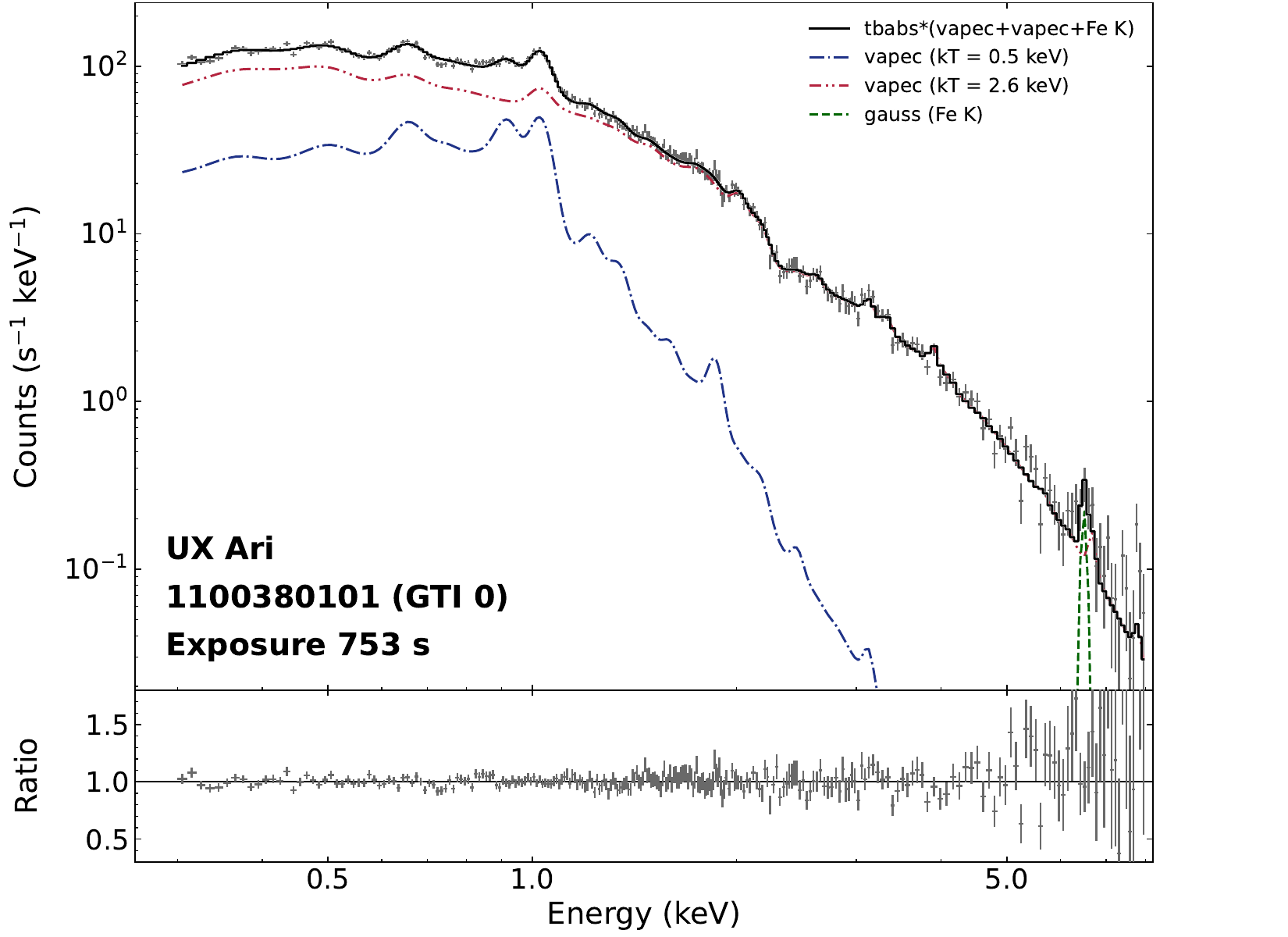}
 \caption{ Background-subtracted 0.3$-$8 keV \textit{NICER} spectrum of Flare U1 (GTI 0 of ObsID 1100380101) fitted with the two-temperature CIE model with interstellar absorption and Fe K$\alpha$ line (\texttt{tbabs*(vapec+vapec+gauss)}; black). The temperature of the two CIE components is 0.5 keV (blue dashed–dotted) and 2.6 keV (red dashed). All best-fit parameters of this spectrum are summarized in Table \ref{tab:1100380101_block0_multi}.}
 \label{fig:1100380101_block0_multi_example}
\end{figure}

\begin{figure}
\centering
 \includegraphics[width=8.5cm]{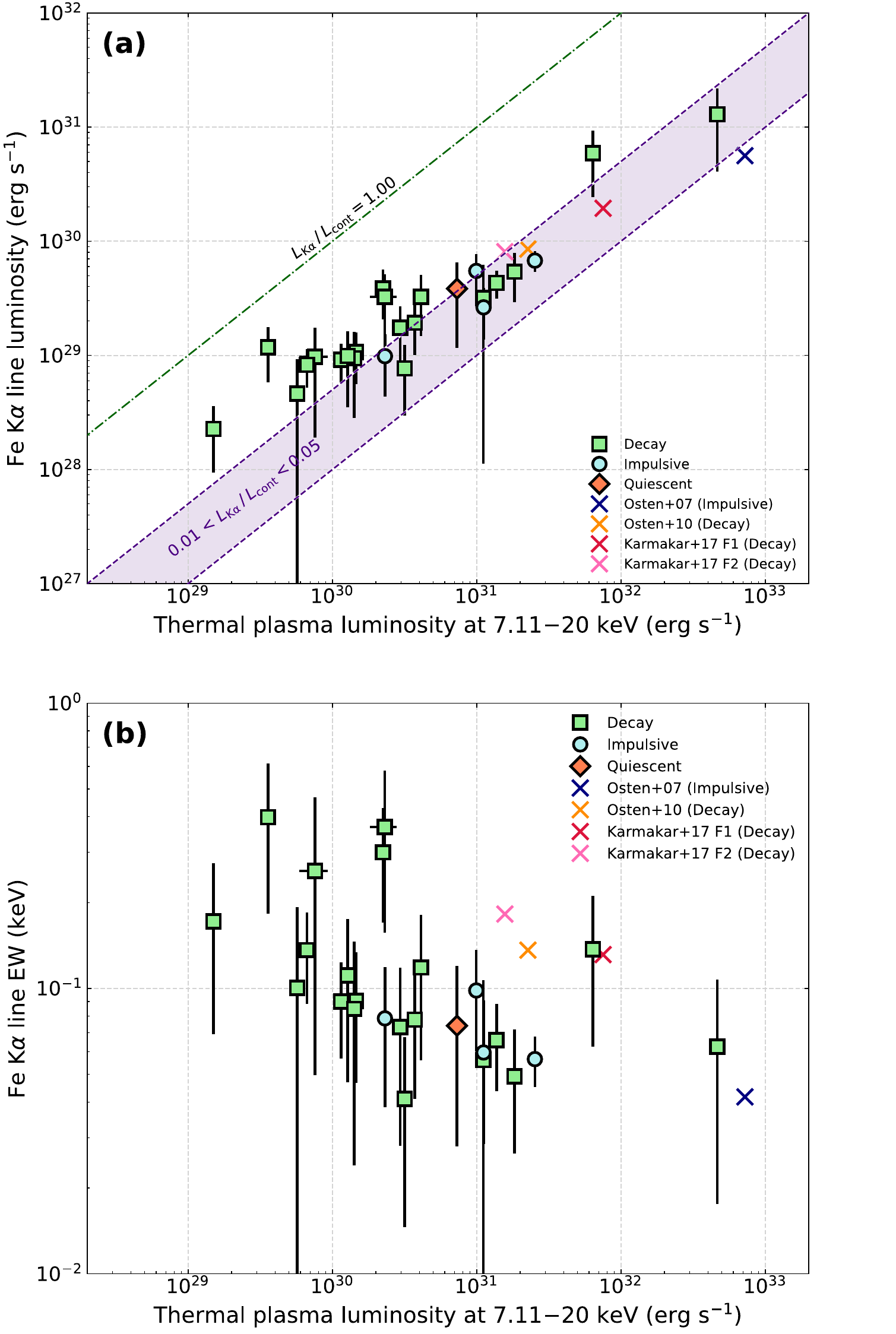}
 \caption{ Same as Figure \ref{fig:HXRluminosity_vs_Kaluminosity_with_lit} and \ref{fig:HXRluminosity_vs_KaEW_with_lit}, but for wide-band (0.3$-$8 keV) fitting results.}
 \label{fig:HXRluminosity_vs_Kaluminosity_multi}
\end{figure}

\begin{table}
\caption{Best-fit parameters of the 1100380101 (GTI 0)spectrum with two temperature collisionally-ionized models. The error ranges correspond to $90\%$ confidence level. Values without errors mean that they are fixed. \vspace{0.5cm}}
\label{tab:1100380101_block0_multi}
\centering
\begin{tabular}{ccc}
\hline
\multicolumn{3}{c}{ObsID: 1100380101 (GTI 0)} \\ \hline 
\multicolumn{2}{c}{\texttt{tbabs}} &  \\ 
\multicolumn{2}{c}{$N_{\mathrm{H}}$ ($10^{19}$ $\mathrm{cm^{-2}}$)} & $8.02\pm 2.31$ \\ \hline
\multicolumn{2}{c}{\texttt{vapec} (Low Temp.)} &  \\
\multicolumn{2}{c}{$kT$ (keV)}  & $0.51 \pm 0.05$ \\
\multicolumn{2}{c}{norm ($10^{-2}$)} & $3.11 \pm 0.77$ \\ \hline
\multicolumn{2}{c}{\texttt{vapec} (High Temp.)} &  \\
\multicolumn{2}{c}{$kT$ (keV)} & $2.61 \pm 0.1$ \\
\multicolumn{2}{c}{norm ($10^{-2}$)} & $13.43 \pm 0.35$ \\ \hline
\multicolumn{2}{c}{\texttt{gauss}} &  \\
\multicolumn{2}{c}{$E_{l}$} & $6.49 \pm 0.06$ \\
\multicolumn{2}{c}{$\sigma$} &  0.0 \\
\multicolumn{2}{c}{$K^{\mathrm{gauss}}$ ($10^{-4}$)} & $1.23 \pm 0.35$ \\ \hline
\multicolumn{2}{c}{$\chi^{2}$ ($d.o.f$)} & 258.04 (254) \\ \hline
\multicolumn{2}{c}{Null hyp. prob.} & 4.18e-01 \\ \hline
\end{tabular}
\end{table}

\renewcommand{\arraystretch}{1.0}
\begin{table*}
	\centering
	\caption{Same as Table \ref{tab:detection_parameters_list}, but for the two-temperature fitting in 0.3$-$8 keV}
	\label{tab:detection_parameters_list_multi}
\begin{tabular}{clccrrrrr}
\hline 
Star                          & Flare     & Obs-ID     & GTI   & $E_{l}$ & $F_{\mathrm{K \alpha}}$ & $L_{\mathrm{K \alpha}}$ & EW$_{\mathrm{K \alpha}}$ & $L_{\mathrm{HXR}}$ \\
                              &           &            &       & (keV)   & ($10^{-4}$ photons cm$^{-2}$ s$^{-1}$)&  ($10^{30}$ erg s$^{-1}$)          & (eV)                     & ($10^{30}$ erg s$^{-1}$)      \\ \hline
\multirow{7}{*}{UX Ari}       & U1$^{\dag}$ & 1100380101 & 0     & $6.49 \pm 0.06$ & $1.23 \pm 0.57$ & $0.39 \pm 0.18$ & $282.8 \pm 127.9$ & $2.25  \pm 0.22$ \\
                              & U2$^{\dag}$ & 1100380106 & all     & $6.49 \pm 0.03$ & $1.73 \pm 0.79$ & $0.54 \pm 0.25$ & $47.2 \pm 25.1$ & $18.19  \pm 0.56$ \\
                              & U3          & 1100380108 & 0+1+2     & $6.42 \pm 0.02$ & $2.16 \pm 0.44$ & $0.68 \pm 0.14$ & $57.4 \pm 11.2$ & $25.3  \pm 0.68$ \\
                              & U4          & 1100380109 & all     & $6.46 \pm 0.09$ & $1.04 \pm 0.57$ & $0.33 \pm 0.18$ & $117.3 \pm 70.0$ & $4.09  \pm 0.3$ \\
                              & U4          & 1100380113 & all     & $6.45 \pm 0.15$ & $0.15 \pm 0.15$ & $0.05 \pm 0.05$ & $105.6 \pm 97.9$ & $0.57  \pm 0.04$ \\
                              & U4          & 1100380118 & all     & $6.51 \pm 0.05$ & $0.38 \pm 0.19$ & $0.12 \pm 0.06$ & $399.4 \pm 202.4$ & $0.36  \pm 0.03$ \\
                              & U5$^{\dag}$ & 1100380127 & 3+4     & $6.6 \pm 0.07$ & $1.04 \pm 0.59$ & $0.33 \pm 0.18$ & $352.6 \pm 207.3$ & $2.29  \pm 0.47$ \\      \hline
\multirow{2}{*}{GT Mus}       & G1          & 1100140102 & 5     & $6.47 \pm 0.06$ & $4.01 \pm 2.35$ & $5.91 \pm 3.46$ & $135.1 \pm 82.8$ & $64.27  \pm 6.83$ \\    
                              & Quiescent    & 1100140108 & all     & $6.43 \pm 0.06$ & $0.26 \pm 0.18$ & $0.38 \pm 0.27$ & $72.2 \pm 46.6$ & $7.41  \pm 0.42$ \\ \hline
\multirow{3}{*}{$\sigma$ Gem} & S1           & 1200040104 & 0     & $6.38 \pm 0.05$ & $-1.43 \pm 0.84$ & $-0.26 \pm 0.15$ & $-26.2 \pm 14.2$ & $25.0  \pm 0.77$ \\
                              & S1           & 1200040104 & 5     & $6.41 \pm 0.12$ & $1.77 \pm 1.7$ & $0.32 \pm 0.31$ & $60.7 \pm 53.1$ & $11.07  \pm 0.9$ \\
                              & S1           & 1200040106 & all     & $6.47 \pm 0.06$ & $0.43 \pm 0.26$ & $0.08 \pm 0.05$ & $38.4 \pm 24.6$ & $3.15  \pm 0.11$ \\     \hline
HD 251108                     & HD1          & 5203530103 & all     & $6.44 \pm 0.06$ & $0.42 \pm 0.28$ & $13.13 \pm 8.89$ & $71.8 \pm 46.6$ & $459.82  \pm 34.74$ \\ \hline 
\multirow{13}{*}{HR1099}      & HR1$^{\dag}$ & 1114010117 & 5     & $6.7 \pm 0.3$ & $0.94 \pm 0.76$ & $0.1 \pm 0.08$ & $271.7 \pm 219.9$ & $0.74  \pm 0.17$ \\
                              & HR2          & 1114010120 & 7     & $6.39 \pm 0.04$ & $1.87 \pm 0.89$ & $0.19 \pm 0.09$ & $68.5 \pm 36.2$ & $3.7  \pm 0.16$ \\
                              & HR2          & 1114010121 & 5     & $6.41 \pm 0.06$ & $1.04 \pm 0.5$ & $0.11 \pm 0.05$ & $90.4 \pm 38.5$ & $1.46  \pm 0.07$ \\
                              & HR2          & 1114010122 & 10+11     & $6.4 \pm 0.03$ & $0.89 \pm 0.34$ & $0.09 \pm 0.03$ & $91.0 \pm 34.3$ & $1.16  \pm 0.04$ \\
                              & HR2          & 1114010123 & 1+2+3     & $6.54 \pm 0.06$ & $0.8 \pm 0.3$ & $0.08 \pm 0.03$ & $141.5 \pm 46.5$ & $0.67  \pm 0.02$ \\
                              & HR3          & 1114010127 & all     & $6.51 \pm 0.07$ & $0.22 \pm 0.13$ & $0.02 \pm 0.01$ & $163.9 \pm 102.2$ & $0.15  \pm 0.01$ \\
                              & HR4$^{\dag}$ & 1114010128 & 2     & $6.49 \pm 0.07$ & $0.92 \pm 0.64$ & $0.09 \pm 0.07$ & $90.6 \pm 59.5$ & $1.42  \pm 0.09$ \\
                              & HR4$^{\dag}$ & 1114010128 & 6     & $6.49 \pm 0.07$ & $0.96 \pm 0.62$ & $0.1 \pm 0.06$ & $108.4 \pm 63.7$ & $1.28  \pm 0.09$ \\
                              & HR5       & 1114010133 & 1     & $6.42 \pm 0.05$ & $2.56 \pm 1.23$ & $0.26 \pm 0.13$ & $59.9 \pm 32.6$ & $11.16  \pm 0.58$ \\
                              & HR5       & 1114010133 & 6     & $6.52 \pm 0.03$ & $5.33 \pm 2.11$ & $0.55 \pm 0.22$ & $98.9 \pm 40.9$ & $9.94  \pm 0.52$ \\
                              & HR5       & 1114010133 & 11     & $6.49 \pm 0.05$ & $1.69 \pm 0.92$ & $0.17 \pm 0.1$ & $77.5 \pm 46.5$ & $2.95  \pm 0.14$ \\
                              & HR6       & 1114010136 & 2     & $6.37 \pm 0.04$ & $0.95 \pm 0.53$ & $0.1 \pm 0.05$ & $75.0 \pm 44.7$ & $2.29  \pm 0.11$ \\
                              & HR7       & 1114010153 & 2+3     & $6.38 \pm 0.02$ & $4.2 \pm 1.15$ & $0.43 \pm 0.12$ & $64.6 \pm 21.0$ & $13.7  \pm 0.41$ \\ \hline
\end{tabular}
\end{table*}
 For comparison with the narrow-band fit (5$-$8 keV) in Section \ref{sec:results}, we also fitted the Fe-K$\alpha$-line detected spectra with the two-temperature CIE model with the gauss at $\sim 6.4$ keV (\texttt{vapec+vpec+gauss}) convolved with interstellar absorption (\texttt{tbabs}) in 0.3$-$8 keV.
 Figure \ref{fig:1100380101_block0_multi_example} and table \ref{tab:1100380101_block0_multi} show the flare spectrum and fitting results of \texttt{tbabs*(vapec+vpaec+gauss)} model as an example.
 Table \ref{tab:detection_parameters_list_multi} summarizes the best-fit parameters of the Fe K$\alpha$ line and $L_{\mathrm{HXR}}$ obtained by the wide-band fitting.
 All figures and tables of the two-temperature spectral analysis are provided as Online material.
 Figure \ref{fig:HXRluminosity_vs_Kaluminosity_multi} shows the $L_{\mathrm{HXR}}-L_{\mathrm{K\alpha}}$ correlation and $L_{\mathrm{HXR}}-\mathrm{EW}_{\mathrm{K\alpha}}$ anti-correlation obtained by the wide-band fitting.

 Most of the best-fit parameters of the Fe K$\alpha$ line obtained by the wide-band fitting (Table \ref{tab:detection_parameters_list_multi}) are consistent within the errors with our previous 5$-$8 keV single-temperature fitting results (Table \ref{tab:detection_parameters_list}).
 Furthermore, the correlation coefficient and powerlaw index between $L_{\mathrm{HXR}}$ and $L_{\mathrm{K\alpha}}$ obtained by the wide-band fitting are $\mu = 0.82 \pm 0.06$ and $L_{\mathrm{K \alpha}} \propto L_{\mathrm{HXR}}^{0.74\pm0.26}$, respectively.
 These values are consistent within the errors with our previous 5$-$8 keV single-temperature fitting results of $\mu = 0.74 \pm 0.08$ and $L_{\mathrm{K \alpha}} \propto L_{\mathrm{HXR}}^{0.86\pm0.46}$.
 When we conduct the 0.3$-$8 keV two-temperature fitting, the $L_{\mathrm{HXR}}$ errors are reduced with smaller errors of the correlation coefficient and powerlaw index.
 The deviation of Flare U4 and HR3 seen in the 5$-$8 keV single-temperature fitting result (Figure \ref{fig:HXRluminosity_vs_Kaluminosity_with_lit}c) is not confirmed in the 0.3$-$8 keV two-temperature fitting result (Figure \ref{fig:HXRluminosity_vs_Kaluminosity_multi}a).
 The correlation coefficient and powerlaw index between $L_{\mathrm{HXR}}$ and $\mathrm{EW}_{\mathrm{K\alpha}}$ obtained by the wide-band fitting are $\mu = -0.48 \pm 0.12$ and $\mathrm{EW}_{\mathrm{K \alpha}} \propto L_{\mathrm{HXR}}^{-0.24\pm0.07}$, respectively, which are also consistent within the errors with our previous 5$-$8 keV single-temperature fitting results of $\mu = -0.44 \pm 0.12$ and $L_{\mathrm{K \alpha}} \propto L_{\mathrm{HXR}}^{-0.27\pm0.10}$.
 Based on these results, we conclude that whether we fit the spectra with the single-temperature model in 5$-$8 keV or two-temperature model in 0.3$-$8 keV does not change significantly the conclusions of this paper.

\clearpage

\section{MPU-divided spectra of $\sigma$ Gem}\label{sec:MPU-divided}

\begin{figure*}
\centering
 \includegraphics[width=14.5cm]{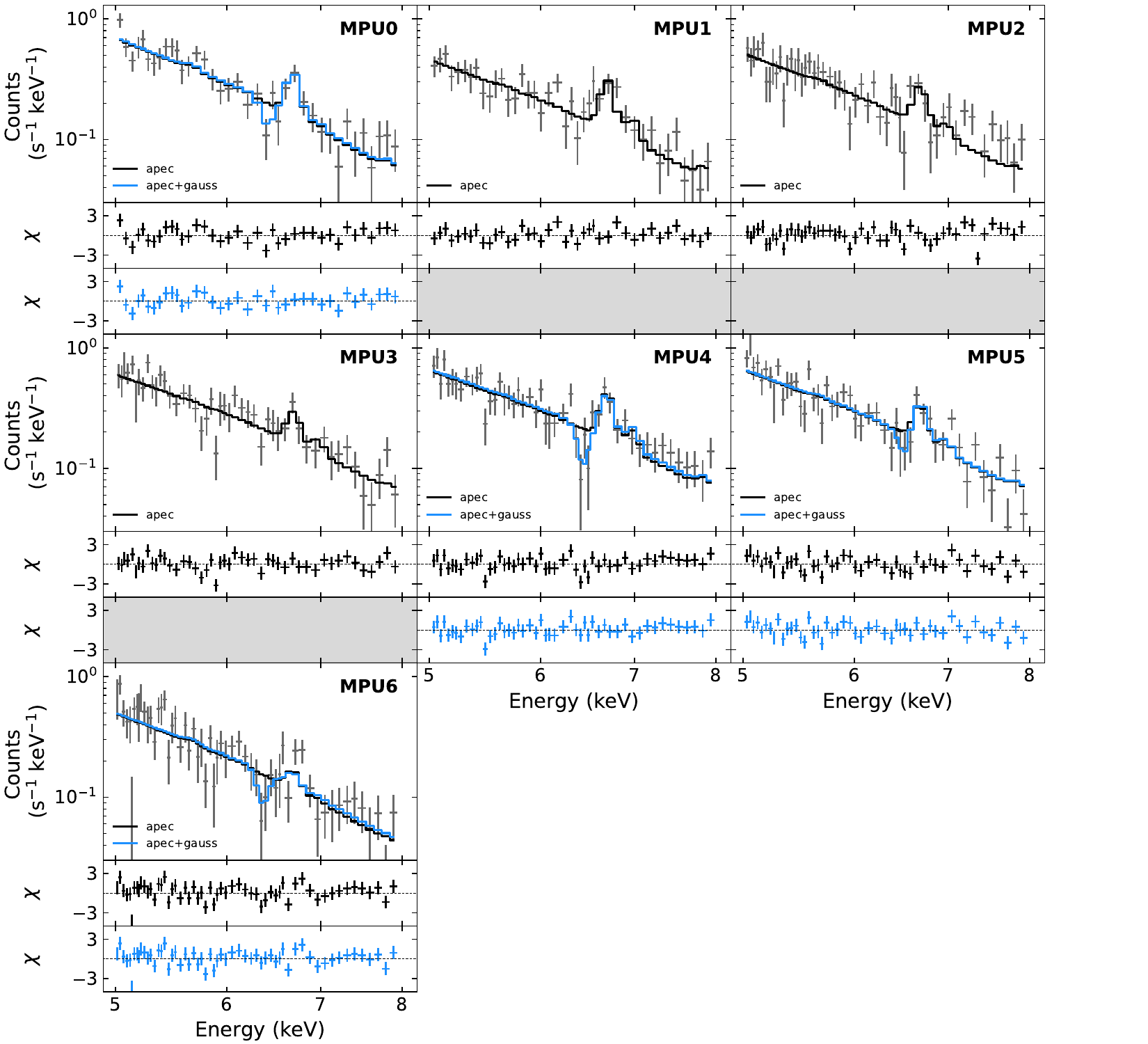}
 \caption{MPU-divided and background-subtracted 5$-$8 keV spectra of $\sigma$ Gem during the decay phase of Flare S1 (GTI 0 of Obs-ID 120040104). Black and blue solid lines correspond to the best-fit model of CIE (\texttt{apec}) and CIE with the absorption line (\texttt{apec+gauss}), respectively.}
 \label{fig:Sigma_Gem_S1_MPU}
\end{figure*}

As discussed in Section \ref{sec:absorption}, the spectrum of $\sigma$ Gem on 2019 February 5 02:15:06$-$02:33:39 during the decay phase of the Flare S1 showed the signature of the absorption line at $\sim 6.4$ keV.
Such absorption line has never been reported neither on the Sun nor other flare stars.
Then, we investigated the possibility that a specific MPU caused the structure of the absorption line due to detector issues.

We created the MPU-divided event files with \texttt{nifpmsel} from that of GTI 0 of Obs-ID 1200040104 extracted by \texttt{niextract-event} (Section \ref{sec:reduction_analyses}).
Then, we extracted the source and background spectra together with response files with \texttt{nicel3-spec} from the MPU-divided event files.
We fitted the MPU-divided spectra with the CIE model (\texttt{apec}) and CIE with the absorption line (\texttt{apec+gauss}) in 5$-$8 keV as the all-MPU spectrum (Figure \ref{fig:Sigma_Gem_S1_main}).

Figure \ref{fig:Sigma_Gem_S1_MPU} shows the MPU-divided spectra with the best-fit models.
The spectra of MPU 0, 4, 5, and 6 can be fitted with the CIE with the absorption line model.
This model is used when the 90\% upper limit of \texttt{gauss} is below 0.
Though the spectra of MPU 1, 2, and 3 cannot be fitted with the CIE with the absorption line model, which is probably due to the statistical fluctuation, they also shows the signature of the absorption line (Figure \ref{fig:Sigma_Gem_S1_MPU}).
Then we conclude that the differences between the spectra of each MPU can be explained by the statistical uncertainty and no specific MPU caused the absorption feature.
This supports the astronomical origin, e.g., the scenario that the geometry of the flare loop caused it (Section \ref{sec:absorption_discussion}).

\section{Signature of P-Cygni profile on HR 1099}\label{sec:PCyg_HR1099}
\begin{figure}
\centering
 \includegraphics[width=8cm]{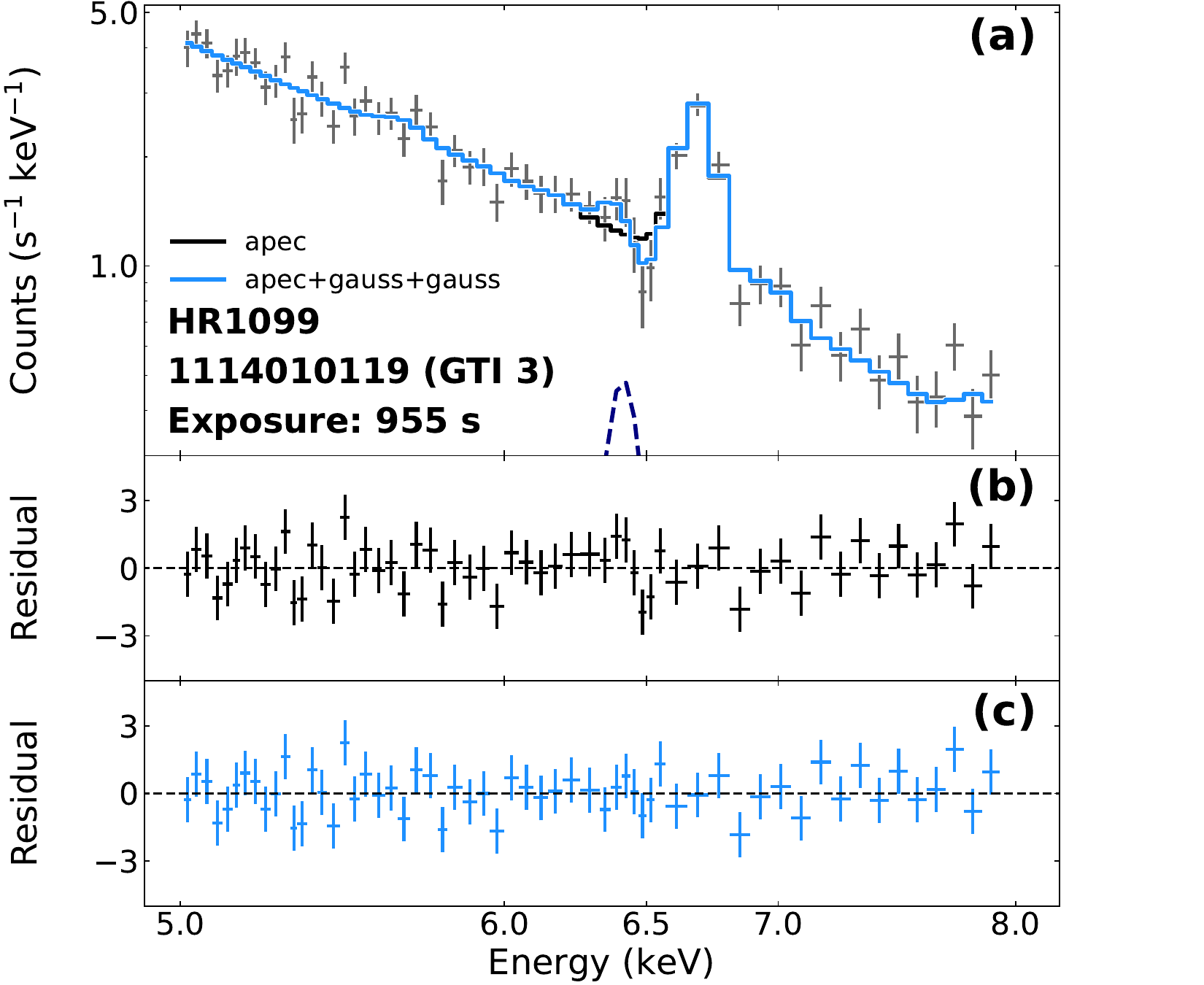}
 \caption{Same as Figure \ref{fig:Sigma_Gem_S1_main}, but for the HR 1099 data of GTI 3 of ObsID 1114010119. All best-fit parameters of this spectrum are summarized in Table \ref{tab:1114010119_block3}.}
 \label{fig:HR1099_HR2_PCyg}
\end{figure}

We found the signature of P-Cygni profile in the HR 1099 data of GTI 3 of ObsID 1114010119 (Figure \ref{fig:HR1099_HR2_PCyg}) during the decay phase of the flare HR2. 
At $\sim 6.4$ keV, both the potential emission and slightly blue-shifted emission component could be seen.
We fitted this spectrum with a CIE model (\texttt{apec}) and a CIE model with the two additional Gaussian components (\texttt{apec+gauss+gauss}).
Table \ref{tab:1114010119_block3} summarizes the best-fit parameters.
The 90 \% upper limit of the absorption component has a positive value (= $2.35 \times 10^{-4}$ photons cm$^{-2}$ s$^{-1}$), which shows the overlap the criteria of emission detection. Because of this, we cannot claim the existence of the absorption component (i.e. the existence of P-Cygni profile) with our criterion (Section \ref{sec:reduction_analyses}), and here we do not discuss more details.

\renewcommand{\arraystretch}{1.3}
\begin{table*}
\caption{Best-fit spectral parameters of \texttt{apec} and \texttt{apec+gauss+gauss} models shown in Figure \ref{fig:HR1099_HR2_PCyg}.}
\begin{center}
\begin{tabular}{cccccc}
\hline 
\multicolumn{6}{c}{1114010119 GTI 3 (During Flare HR2 on HR 1099)}  \\ \hline 
\multicolumn{3}{c|}{Without the additional Gaussiann}  & \multicolumn{3}{c}{With the additional Gaussiann} \\ \hline
\multirow{3}{*}{\texttt{apec}}                          & $kT$ (keV) / $T$ (MK)        & \multicolumn{1}{c|}{$4.11^{+0.58}_{-0.43}$ / $47.7^{+6.7}_{-5.0}$} & \multirow{3}{*}{\texttt{apec}}    & $kT$ (keV) / $T$ (MK)        & $4.10^{+0.53}_{-0.43}$ /  $47.6^{+6.2}_{-5.0}$   \\
                                                        & $v \: (\mathrm{km \: s^{-1}})$               & \multicolumn{1}{c|}{$0.00$ (fix)} &                                   & $v \: (\mathrm{km \: s^{-1}})$ & $0.00$ (fix)   \\
                                                        & $K^{\mathrm{apec}}$              & \multicolumn{1}{c|}{$0.67^{+0.10}_{-0.09}$} &                                   & $K^{\mathrm{apec}}$ & $0.67^{+0.11}_{-0.05}$    \\ \cline{4-6}
\multirow{3}{*}{---}                          & --- & --- & \multirow{3}{*}{\texttt{gauss}}    & $E_{l}$ (keV) &   $6.41^{+0.08}_{-0.10}$ \\
                                                    & --- & --- &                                  & $\sigma$(keV) &   $0.00$ (fix)  \\
                                                    & --- & ---  &                       & $K^{\mathrm{gauss}}$ ($10^{-4}$) &  $2.36^{+34.9}_{-1.98}$  \\ \cline{4-6}
\multirow{3}{*}{---}                          & --- & --- & \multirow{3}{*}{\texttt{gauss}}    & $E_{l}$ (keV) &   $6.46^{+0.11}_{-0.05}$ \\
                                                    & --- & --- &                                  & $\sigma$(keV) &   $0.00$ (fix)  \\
                                                    & --- & ---  &                       & $K^{\mathrm{gauss}}$ ($10^{-4}$) &  $-2.18^{+4.53}_{-37.6}$  \\ \hline
\multicolumn{2}{c}{$\chi^{2}$ (d.o.f.)}                                     & \multicolumn{1}{c|}{57 (56)} & \multicolumn{2}{c}{$\chi^{2}$ (d.o.f.)} & 50 (52) \\
\multicolumn{2}{c}{Null hyp. prob.} & \multicolumn{1}{c|}{0.45} & \multicolumn{2}{c}{Null hyp. prob.} &   0.54  \\ \hline 
\end{tabular}
\end{center}
\label{tab:1114010119_block3}
\end{table*}

%TC:endignore
%%%%%%%%%%%%%%%%%%%%%%%%%%%%%%%%%%%%%%%%%%%%%%%%%%

% Don't change these lines
\bsp	% typesetting comment
\label{lastpage}
\end{document}